\documentclass[10pt,aps,prd,twocolumn,notitlepage,bibnotes,longbibliography,
floatfix,showpacs,citeautoscript,superscriptaddress]{revtex4-2} 

\usepackage[english]{babel}		
\makeatletter\AtBeginDocument{\let\@elt\relax}\makeatother 
\usepackage{amsmath}		
\usepackage{amssymb}		
\usepackage{mathtools}		
\usepackage{bbm}			

\usepackage{array}			
\usepackage{graphicx}		
\usepackage{graphics}		
\DeclareGraphicsExtensions{.jpg,.jpeg,.png,.pdf,.eps}	

\usepackage[pdfstartview=FitV, pdfusetitle,
			bookmarks=true,bookmarksnumbered=false,bookmarksopen=false,
			breaklinks=true,pdfborder={0 0 0},backref=false,colorlinks=true, 
			linkcolor=blue, citecolor=blue, urlcolor=blue]{hyperref}

\renewcommand{\theequation}{\arabic{equation}}

\makeatletter
\newcommand*{\Equation}{\@ifstar\sEquation\oEquation}
\newcommand{\sEquation}[1]{\begin{equation*}#1\end{equation*}}
\newcommand{\oEquation}[2]{  \begin{equation}\label{#1}#2\end{equation} }
\makeatother
\newcommand{\Align}[2]{\begin{align}\label{#1}#2\end{align}}
\newcommand{\SubAlign}[2]{\begin{subequations}\label{#1}
						  \begin{align}#2\end{align}\end{subequations}}

\newcommand{\bs}{\boldsymbol}

\newcommand{\Figref}[1]{Fig.~\ref{#1}}

\newcommand{\Eqref}[1]{\eqref{#1}}
\newcommand{\eq}[1]{(\ref{#1})}

\newcommand{\eg}{{\it e.g.~}}

\newcommand{\ie}{{\it i.e.~}}

\newcommand{\groupZ}[1]{\mathbb{Z}_{#1}} 			

\newcommand{\Exp}[1]{\text{e}^{#1}}

\renewcommand\Re{\mathrm{Re}}
\renewcommand\Im{\mathrm{Im}}

\newcommand{\x}{\bs x}
\newcommand{\Grad}{{\bs\nabla}}
\newcommand{\Div}{{\bs\nabla}\cdot}
\newcommand{\Curl}{{\bs\nabla}\times}

\newcommand{\tx}{\tilde{\bs x}}
\newcommand{\tGrad}{\tilde{\bs \nabla}}
\newcommand{\tDiv} {\tilde{\bs \nabla}\!\cdot\!}
\newcommand{\tCurl}{\tilde{\bs \nabla}\!\times\!}

\newcommand{\F}{\mathcal{F}}
\newcommand{\tF}{\tilde{\mathcal{F}}}

\newcommand{\D}{{\bs D}}
\newcommand{\cD}{{\bs {\mathcal{D}}}}

\newcommand{\cL}{{\bs{\mathcal{L}}}}
\newcommand{\Q}{\bs{\mathcal{Q}}}
\newcommand{\A}{{\bs A}}

\newcommand{\B}{{\bs B}}

\newcommand{\vH}{{\bs H}}
\newcommand{\M}{{\bs M}}
\newcommand{\J}{{\bs J}}
\newcommand{\Jh}{{\bs{\hat J}}}
\renewcommand{\j}{{\bs j}}
\newcommand{\jh}{{\bs{\hat \jmath}}}
\newcommand{\vv}{{\bs v}}

\allowdisplaybreaks
\begin{document}
\title{Counterpart of the Chandrasekhar-Kendall state in noncentrosymmetric superconductors}

\author{Julien Garaud}
\email{garaud.phys@gmail.com}
\affiliation{Institut Denis Poisson CNRS/UMR 7013, 
			 Universit\'e de Tours, 37200 France}

\author{Anatolii Korneev}
\affiliation{Pacific Quantum Center, 
			 Far Eastern Federal University, Sukhanova 8, Vladivostok, 690950, 
			 Russia}

\author{Albert Samoilenka}
\affiliation{Department of Physics, Royal Institute of Technology, 
			SE-106 91 Stockholm, Sweden}

\author{Alexander Molochkov}
\affiliation{Pacific Quantum Center, 
			 Far Eastern Federal University, Sukhanova 8, Vladivostok, 690950, 
			 Russia}

\author{Egor Babaev}
\affiliation{Department of Physics, Royal Institute of Technology, 
			SE-106 91 Stockholm, Sweden}

\author{Maxim Chernodub}
\email{maxim.chernodub@idpoisson.fr}
\affiliation{Institut Denis Poisson CNRS/UMR 7013, 
			 Universit\'e de Tours, 37200 France}

\date{\today}

\begin{abstract}

We demonstrate that superconductors with broken inversion symmetry support a family 
of stable, spatially localized configurations of the self-knotted magnetic field. 
These solutions, that we term ``toroflux,'' are the superconducting 
counterparts of the Chandrasekhar-Kendall states (spheromaks) that appear in highly 
conducting, force-free astrophysical and nuclear-fusion plasmas. 
The superconducting torofluxes are solutions of superconducting models, in the presence 
of a parity-breaking Lifshitz invariant associated with the $O$ point-group symmetry. 
These solutions are characterized by a non-vanishing helicity of the magnetic field, 
and also by a toroidal dipole moment of the magnetic field.
We demonstrate that a magnetic dipole or a ferromagnetic inclusion in the bulk 
of a noncentrosymmetric superconductor sources finite-energy toroflux solutions.
\end{abstract}

\maketitle

\section{Introduction}

Ordinary type-2 superconductors expel weak magnetic fields due to the Meissner effect, 
while at elevated fields the magnetic flux penetrates in the form of a lattice or a liquid 
of  vortices (see, \eg, \cite{Tinkham}).  Moreover, quantum or thermal fluctuations 
can induce closed loops of such quantum vortices. Because of the vortex string tension, 
these loops are unstable, and eventually decay. Thus, apart from certain cases demonstrated 
in multicomponent systems that allow different topology \cite{Rybakov.Garaud.ea-2019} bulk 
superconductors do not feature stable, localized configurations of the magnetic field 
(in three dimensions).
In this paper, we demonstrate that bulk noncentrosymmetric superconductors feature 
a class of localized, impurity-induced, configurations of a knotted magnetic field. 
We coin these solutions ``toroflux,'' since the geometry of their current and flux lines 
resemble a popular toroflux toy \cite{Wikipedia}.

Noncentrosymmetric superconductors, that is superconductors whose crystal lattices 
lack inversion symmetry, have attracted significant attention from both theoretical 
\cite{Bulaevskii.Guseinov.ea-1976,Levitov.Nazarov.ea-1985,Mineev.Samokhin-1994,
Edelstein-1996,Agterberg-2003,Samokhin-2004} and experimental~\cite{Bauer.Hilscher.ea-2004,
Samokhin.Zijlstra.ea-2004,Yuan.Agterberg.ea-2006,Cameron.Yerin.ea-2019,
Khasanov.Gupta.ea-2020} communities. 
A key property of a noncentrosymmetric crystal is that it cannot be superimposed on its 
spatially inverted image with the help of spatial translations. The crystal thus breaks 
explicitly the parity inversion group. Since the superconducting order parameter captures 
the parity-breaking properties of the underlying ionic lattice, the noncentrosymmetric 
superconductors constitute a class of exotic systems that spontaneously breaks 
a continuous symmetry, in a parity-violating medium (see, \eg, 
Refs.~\cite{Bauer.Sigrist,Yip-2014,Smidman.Salamon.ea-2017} for detailed reviews). 
Ginzburg-Landau free energies of noncentrosymmetric superconductors include contributions 
that are linear in the magnetic field and in the gradients of the superconducting order 
parameter: $\propto k_{ij}B_i\Im(\psi^*D_j\psi)$. Here $\D$ is the gauge derivative of 
the order parameter $\psi$, and $k_{ij}$ are coefficients that depend on the crystal 
symmetry. In this work, we consider a particular class of noncentrosymmetric superconductors 
with chiral octahedral $O$ symmetry.

Parity-breaking superconducting systems feature several distinctive properties: 
they generate unusual magnetoelectric transport phenomena, exhibit a correlation between 
supercurrents and electron spin polarizations, lead to the emergence of helical states, 
and host, in the background of the magnetic field, the vortex lattices with exotic spatial 
structure~\cite{Bauer.Sigrist,Yip-2014,Smidman.Salamon.ea-2017,Kashyap.Agterberg-2013}. 
Notably, vortices in these superconducting materials can exhibit an inversion of the 
magnetic field at a certain distance from the vortex core 
\cite{Garaud.Chernodub.ea-2020,Samoilenka.Babaev-2020}. This property leads to non-monotonic 
inter-vortex forces and thus to the formation of vortex-vortex bound states, vortex 
clusters, and nontrivial bound states at the boundary of the 
sample~\cite{Garaud.Chernodub.ea-2020,Samoilenka.Babaev-2020}. 
The parity breaking in noncentrosymmetric superconductors can also modify the Josephson 
effect with an unconventional, phase-shifted relation for the Josephson 
current~\cite{Buzdin-2008,Konschelle.Buzdin-2009}. Linked by a uniaxial ferromagnet, the 
unconventional Josephson junction was suggested to serve as an element of a qubit with 
a simple and presumably robust architecture~\cite{Chernodub.Garaud.ea-2022}.

The toroflux solutions that we find in this paper, are the counterparts of the 
Chandrasekhar-Kendall states~\cite{Chandrasekhar.Kendall-1957}, in the context of 
noncentrosymmetric superconductors. The Chandrasekhar-Kendall states are the 
divergence-free eigenvectors of the curl operator that determine the minimum-energy 
equilibrium configurations in magnetohydrodynamics of highly conducting plasmas. 
These states appear in various physical contexts, ranging from astrophysical 
plasmas~\cite{Chandrasekhar.Kendall-1957} to the nuclear fusion theory~\cite{Taylor-1974}. 
In the latter case, the Chandrasekhar-Kendall eigenvectors are also known as Taylor 
states~\cite{Taylor-1974}, which represent the relaxed minimum energy states of a plasma 
in a spheromak device (\ie, inside a spherical shell that confines the plasma) 
\cite{Rosenbluth.Bussac-1979,Taylor-1986}. The principal difference between the toroflux 
state in parity-broken superconductors and the Chandrasekhar-Kendall state in a 
conducting plasma is that the toroflux are strongly localized configurations. The spatial 
localization of both the magnetic field and the supercurrent of the toroflux originates 
from the Meissner effect.

Our torofluxes are eigenstates of the London equations for a noncentrosymmetric 
superconducting material. Labeled by their orbital ($0<l<\infty$) and magnetic 
($-l\leqslant m\leqslant+l$) quantum numbers, there are infinitely many $(l,m)$ toroflux 
modes, for a given value of the parity-breaking parameter. 
All of the toroflux modes have an intrinsic divergence at the origin, and therefore 
they require a regularization at the core of the solutions. We demonstrate that each 
divergent mode is regularized by (pointlike) magnetic multipole sources. The case of 
a pointlike magnetic dipole is of particular physical relevance, as it corresponds to 
magnetic impurities inside a noncentrosymmetric superconductor. We argue that such 
magnetic impurities systematically induce an $(l,m)=(1,0)$ toroflux mode. 

The superconducting toroflux solutions found in this paper share some similarities with 
knotted field configurations that appear in many areas of physics, including particle physics 
\cite{Faddeev.Niemi-1997}, condensed matter~\cite{Babaev-2002b,Rybakov.Garaud.ea-2019,
Sutcliffe-2017,Rybakov.Kiselev.ea-2022}, and the classical field theory 
\cite{Radu.Volkov-2008,Battye.Sutcliffe-1998}. Knotted electromagnetic field configurations 
were also suggested to play a role in the chirally imbalanced quark-gluon plasmas
\cite{Chernodub-2010,Tuchin-2015,Hirono.Kharzeev.ea-2015,Tuchin-2016,Xia.Qin.ea-2016}.

The paper is organized as follows. In Sec.~\ref{Sec:Theory}, we introduce the 
Ginzburg-Landau theory for parity-breaking superconductors and derive the corresponding 
classical equations in the London limit. In Sec.~\ref{Sec:Compact:solution}, we express 
the London equation in terms of a force-free field and discuss localized solutions for the 
magnetic field and electric currents, using the basis of vector spherical harmonics. 
There, we also determine the energy and helicity densities for the infinite tower 
of toroflux states. We further demonstrate that in the London limit, the total energy 
of the solution diverges in its core. 
Next, in Sec.~\ref{Sec:Inclusion}, we show that a ferromagnetic inclusion regularizes 
the singular behavior of the solution, serving, at the same time, as a source for a 
finite-energy superconducting toroflux. 
Finally, in Sec.~\ref{Sec:Dipole} we investigate the case where the inclusion is 
a ferromagnetic dipole. There, we explicitly construct the toroflux solutions sourced 
by such an impurity. We discuss their properties and, in particular, the influence of 
the parity-breaking parameter on the structure, energy, and helicity of the toroflux 
solutions.
Our conclusions are presented in the last section. 

\section{Theoretical framework}
\label{Sec:Theory}

\subsection{Parity-broken formulation}

We consider a class of isotropic noncentrosymmetric superconductors that are invariant 
under spatial rotations while possessing, at the same time, an explicitly broken discrete 
group of spatial inversions. The macroscopic physics of these materials may be described 
within the Ginzburg-Landau theory supplemented with the Lifshitz term of the simplest form 
$\j\cdot\B$ that directly couples the magnetic field $\B$ to a current $\j$ expressed via 
the superconducting order parameter $\psi$ (for a review, see 
Refs.~\cite{Bauer.Sigrist,Agterberg-2012}). 
This particular structure of the Lifshitz term describes a class of the noncentrosymmetric 
superconductors with an $O$-point group symmetry such as, for example, Li$_2$Pt$_3$B 
\cite{Badica.Kondo.ea-2005,Yuan.Agterberg.ea-2006}, Mo$_3$Al$_2$C 
\cite{Karki.Xiong.ea-2010,Bauer.Rogl.ea-2010}, and PtSbS \cite{Mizutani.Okamoto.ea-2019}.

In the vicinity of the superconducting critical temperature, the density $\F$ of the 
Ginzburg-Landau free energy $F=\int d^3x \F$ can we written as follows:
\SubAlign{Eq:FreeEnergy}{
 \F&= \frac{\B^2}{8\pi}+
 \frac{k}{2}\sum_{a=\pm}\big|\cD_a\psi\big|^2
	 +\frac{\beta}{2}(|\psi|^2-\psi_0^2)^2 \,,\\
	 &~~~\text{where}~~\cD_\pm:=\Grad-ie\A+ie\varkappa_\pm\B 	 \,.
}
The single-component order parameter $\psi=|\psi|\Exp{i\varphi}$ stands for the density 
of Cooper pairs. The gauge derivative $\cD$ couples the scalar field $\psi$ to the vector 
potential $\A$ and the magnetic field $\B=\Curl\A$. The coefficients of the Ginzburg-Landau 
model~\eq{Eq:FreeEnergy}, including the parity-breaking couplings $\varkappa_\pm=\chi\pm\nu$, 
can be expressed in terms of the parameters of the microscopic 
model~\cite{Samoilenka.Babaev-2020} (see also~\cite{Mineev.Samokhin-2008}). 
In the microscopic single-particle Hamiltonian, the parity-odd terms originate from 
the antisymmetric spin-orbit couplings ${\bs g}_{\bs k}\cdot {\bs \sigma}$ with $
{\bs g}_{\bs k} = - {\bs g}_{- \bs k}$ and the Pauli matrices $\bs\sigma$ acting on 
the spin space~\cite{Kaur.Agterberg.ea-2005}. The parity-odd couplings $\varkappa_\pm$, 
allowed by the parity-broken nature of the noncentrosymmetric superconductors, are 
nonvanishing but have parametrically small values~\cite{Bauer.Sigrist}. Throughout the 
paper, we use the units $\hbar = c = 1$.

The last term in the gauge derivative $\cD$ that appear in the free 
energy~\eq{Eq:FreeEnergy} corresponds to the Lifshitz invariant associated with the $O$ 
point-group symmetry. This term reflects the breaking of the parity symmetry in the system. 
Indeed, under the parity inversion, $P(\x)=-\x$, the magnetic field transforms as a 
parity-even quantity $P(\B)=\B$ while the other terms in the derivative transform as 
vectors, $P(\cD_\pm)=-\cD_\pm+2ie\varkappa_\pm\B$. Thus, the presence of the last term 
in the derivative makes the free energy density \eq{Eq:FreeEnergy} noninvariant under 
the parity inversion: $P(\F)=\F-2ek(\varkappa_++\varkappa_-)\B\cdot\Im(\psi^*\D\psi)$.

The physical length scales of the theory, namely, the coherence length $\xi$ and the London 
penetration depth $\lambda_L$, are determined by the coefficients of the Ginzburg-Landau 
model as
\SubAlign{Eq:LengthScales}{
\lambda_L&=\lambda_0\sqrt{1+\frac{\varkappa_+^2+\varkappa_-^2}{2\lambda_0^2}}
\,,~\text{where }~~
\lambda_0^2=\frac{1}{8\pi ke^2\psi_0^2} \,, \\
 \xi^2&=\frac{k}{2\beta\psi_0^2}  
\,,}
respectively. Note that in noncentrosymmetric superconductors, an externally applied 
magnetic field does not decay in a simple monotonic way; for a detailed discussion of 
how a counterpart of the London's penetration length is defined in such a case see, 
\eg, Refs.~\cite{Garaud.Chernodub.ea-2020,Samoilenka.Babaev-2020}.

The variation of the free energy \Eqref{Eq:FreeEnergy} with respect to the field $\psi^*$ 
yields the Ginzburg-Landau equation for the superconducting condensate
\Equation{Eq:EOM:GL}{
k\sum_{a=\pm}\cD_a\cD_a\psi=2\beta(|\psi^2|-\psi_0^2)\psi\,,
}
while the variation with respect to the vector potential $\A$ determines the 
Amp\`ere-Maxwell equation
\Align{Eq:EOM:AM}{
&\Curl\Big(\frac{\B}{4\pi}+ke\sum_{a=\pm}\varkappa_a\J_a\Big)=ke\sum_{a=\pm}\J_a\,,\\
&~~~~~\text{where}~~\J_a=\Im(\psi^*\cD_a\psi) \nonumber \,.
}

The structure of the magnetic field lines can be conveniently characterized by 
the magnetic helicity:
\Equation{Eq:Helicity:0}{
{\cal H}=\int\A\cdot\B	\,.
}
It can indeed serve as a measure of the entanglement of the magnetic field lines in knotted 
configurations of the magnetic field~\cite{Berger-1999}. The magnetic helicity is widely 
used in ordinary electrically conducting plasmas described by ideal magnetohydrodynamics, 
where it is a conserved quantity modulo energy-costly reconnections of the magnetic field lines 
\cite{Taylor-1986}.

\subsection{London-limit}

The kinetic term of the free energy \Eqref{Eq:FreeEnergy}, can be expanded into a sum of 
gauge-invariant terms:
\Align{Eq:FreeEnergy:Kinetic}{
 &\frac{1}{2}\sum_{a=\pm}\big|\cD_a\psi\big|^2 =|\D\psi|^2 + \chi \j\cdot\B
	+e^2(\chi^2+\nu^2)|\psi|^2\B^2\,, \nonumber\\
 &\text{where}~~\D:=\Grad-ie\A	~\text{and}~
 \j=2e|\psi|^2(\Grad\varphi-e\A)
 \,.
}
In the London limit
the superconducting density is a spatially 
uniform quantity, $|\psi|=\psi_0$, and the free energy reads as follows~\footnote{Note 
that despite the free energies of parity-breaking Ginzburg-Landau models are not always 
bounded from below (see Ref.~\cite{Samoilenka.Babaev-2020} and also a remark in the appendix 
of Ref.~\cite{Garaud.Chernodub.ea-2020}), their London limits converge to the same 
model~\eq{Eq:FreeEnergy:London:1}.}:
\Align{Eq:FreeEnergy:London:1}{ 
 \F_L&= k\lambda_L^2e^2\psi_0^2\left\lbrace\B^2+\jh^2+2\Gamma\jh\cdot\B  \right\rbrace\,,
 \\
 \text{where}~\jh&=\frac{\j}{2\lambda_Le^2\psi_0^2}\,,~~
 \Gamma=\frac{\chi}{\lambda_L} \,,~~
 \text{and}~0\leqslant\Gamma\leqslant1 
 \,.  \nonumber
}
Importantly, the dimensionless parameter $\Gamma$ quantifies the importance of the parity 
breaking. At $\Gamma=0$, the material is thus centrosymmetric. The second London equation 
that relates the magnetic field $\bs B$ and the current~\eq{Eq:FreeEnergy:Kinetic} 
$\j=2e\psi_0^2(\Grad\varphi-e\A)$ takes the following form:
\Equation{Eq:London:B}{
  \B =\Phi_0\vv-\tCurl\jh \,.
}
Here $\vv=\frac{1}{2\pi}\Curl\Grad\varphi$ is the density of vortex field that accounts 
for the phase singularities, and $\Phi_0=2\pi/e$ is the superconducting flux quantum. 
In the dimensionless units used here $\tx={\x}/{\lambda_L}$ and $\tGrad=\lambda_L\Grad$, 
the Amp\`ere-Maxwell equation \Eqref{Eq:EOM:AM} reads as
\Align{Eq:AM:London}{
&\tCurl\vH=\tCurl(\B-4\pi\M)=\Jh\,,		\\
\text{where}~~&\vH=\B+\Gamma\jh	\,,~~\Jh=\jh+\Gamma\B	\,,
~~\text{and}~~
\M=-\frac{\Gamma\jh}{4\pi} \,,\nonumber
}
are, respectively, the (dimensionless) magnetic field, the total current, 
and the magnetization.

Introducing the complex quantity $\eta=\Gamma+i\sqrt{1-\Gamma^2}$, the free-energy 
density \Eqref{Eq:FreeEnergy:London:1} in the London limit can further be rewritten as
\Equation{Eq:FreeEnergy:London:2}{
 \tF_L:=\frac{\F_L}{k\lambda_L^2e^2\psi_0^2}=(\B+\eta\jh)(\B+\eta^*\jh)
 \,. 
}

The constant density approximation, together with the expression for the magnetic 
field \Eqref{Eq:London:B}, thus yields the dimensionless free energy:
\Equation{Eq:FreeEnergy:London}{
 \tF_L= (\cL^*\jh-\Phi_0\vv)\!\cdot\!(\cL\jh-\Phi_0\vv) \,.
}
Here, for shorthand notation, we introduce the operator $\cL\jh=\tCurl\jh-\eta\jh$.
The London equation for the current $\jh$, obtained as the Euler-Lagrange equation 
by varying the free energy \Eqref{Eq:FreeEnergy:London} with respect to $\jh$, reads as 
\Equation{Eq:London:j}{
 \cL\cL^*\jh =\Phi_0\Re\left[\cL^*\vv \right]	\,.
}
Note that, the source field $\vv$ is not a regular function but a distribution that is zero 
almost everywhere, except for a set of phase singularities identified with positions of 
vortices. Since we are interested in vortex-free configurations, the source term associated 
with the vortex fields is, from now on, set to zero $\vv=0$.

As we demonstrate below, the London equation \Eqref{Eq:London:j} can be seen as a complex, 
force-free equation whose solution corresponds to the eigenfunctions of the curl operator 
with complex eigenvalues. The general axisymmetric eigenfunctions of the curl operator can, 
for example, be found by using the Chandrasekhar-Kendall toroidal-poloidal decomposition 
\cite{Chandrasekhar-1956,Chandrasekhar.Kendall-1957}. Below, we will express the solutions 
differently, using the basis of vector spherical harmonics.

\section{Localized force-free solutions}
\label{Sec:Compact:solution}

We are interested in finding the spatially localized solutions of the London equation 
\Eqref{Eq:London:j}. This equation can be simplified by introducing a complex, 
force-free vector field $\Q$ that satisfies the force-free equation:
\Equation{Eq:London:FF:0}{
\cL\Q=0\,. 
}
Hence, in the absence of a source term, the London equation implies that 
\Equation{Eq:London:FF}{
\cL^*\jh=i\Im(\eta)\Q\,,
}
where $\Q$ obeys the force-free equation \Eqref{Eq:London:FF:0}. The definition \Eqref{Eq:London:FF} relates the physical magnetic fields and the electric current to the force-free field $\Q$ as 
\SubAlign{Eq:London:FF:Physical}{
\jh&=\Re\Q		\,,~&\J =\sqrt{1-\Gamma^2}\Im(\eta\Q)\,,		\\
\B&=-\Re(\eta\Q)\,,~&\vH=\sqrt{1-\Gamma^2}\Im(\Q)\,.
}
It is convenient to represent the solutions $\Q$ of the force-free equation 
\Eqref{Eq:London:FF:0} in the basis of the vector spherical harmonics 
${\bs Z}_{lm} =({\bs Y}_{lm},{\bs \Psi}_{lm},{\bs \Phi}_{lm})$: 
\Equation{Eq:VSH:decomposition}{
\Q(\x) = \sum_{l=0}^\infty \sum_{m=-l}^{+l}\left(~~~~
\sum_{\mathclap{{\bs Z} = {\bs Y}, {\bs \Psi}, {\bs \Phi}} }
	Q_{lm}^{\bs Z\!}(r) \, {\bs Z}_{lm}({\hat {\bs r}}) \right) \,,
}
where the harmonics ${\bs Z}_{lm}$ and the corresponding radial functions 
$Q_{lm}^{\bs Z\!}(r)$ are labeled by the integer-valued quantum number of the angular 
momentum $l=0,1,2,\dots$ and its projection on the $z$ axis, $m\equiv m_z\in\groupZ{}$ with 
$ -l \leqslant m \leqslant l$. The angular coordinates are encoded in the unit vector 
${\hat {\bs r}} \equiv {\bs r}/r$. The vector spherical harmonics are defined, in the 
parametrization of Ref.~\cite{Barrera.Estevez.ea-1985}, via their scalar counterpart 
$Y_{lm} ({\hat {\bs r}})$ as 
\SubAlign{Eq:VSH}{
{\bs Y}_{lm}(   {\hat {\bs r}}) & = Y_{lm} ({\hat {\bs r}}) {\hat {\bs r}}\,, \\
{\bs \Psi}_{lm}({\hat {\bs r}}) & = r {\bs \nabla} Y_{lm} ({\hat {\bs r}})\,,\\
{\bs \Phi}_{lm}({\hat {\bs r}}) & = {\bs r} \times {\bs \nabla} Y_{lm} ({\hat {\bs r}})\,.
}

Given the decomposition \Eqref{Eq:VSH:decomposition}, the force-free equation 
\Eqref{Eq:London:FF:0} yields a set of differential equations whose solutions that are 
bounded at infinity are as follows (see details in Appendix~\ref{App:Derivation}):
\Align{Eq:FF:solution}{
& Q_{lm}^{\bs \Phi\!} = c_{lm} h_l^{(1)}(\eta r)\,,~~ 
  Q_{lm}^{\bs    Y\!} = -c_{lm}\frac{l(l+1)}{\eta r}h_l^{(1)}(\eta r) 	 \nonumber\,,\\
& Q_{lm}^{\bs \Psi\!} = -c_{lm} \left(\frac{l+1}{\eta r}h_l^{(1)}(\eta r) 
-h_{l+1}^{(1)}(\eta r)  \right) \,.
}
Here $c_{lm}$ is an arbitrary complex constant, and $h_l^{(1)}(z)$ is the spherical 
Hankel function of the first kind. Using the relations between the physical 
fields and the force-free field \Eqref{Eq:London:FF:Physical}, the total London free 
energy \Eqref{Eq:FreeEnergy:London} can be written in the basis of the vector spherical 
harmonics as 
\Equation{Eq:Energy}{
\tilde{F}=(1-\Gamma^2)\sum_{l=0}^\infty\sum_{m=-l}^l\int_0^\infty r^2dr
~\sum_{\mathclap{{\bs Z} = {\bs Y}, {\bs \Psi}, {\bs \Phi}} }
~w_{lm}^{\bs Z\!}\Big|Q_{lm}^{\bs Z\!}\Big|^2\,,
}
where $w_{lm}^{\bs Y\!}=1$ and $w_{lm}^{\bs \Phi\!}=w_{lm}^{\bs \Psi\!}=l(l+1)$. 
Here, the angular degrees of freedom have been integrated out using the orthogonality 
properties of the spherical harmonics (see Appendix~\ref{App:VSH}). Note that the 
dimensionless energy \Eqref{Eq:Energy} is related to the total free energy as 
$F=k\lambda_L^5e^2\psi_0^2\tilde{F}$.

According to the definitions of the free energy \Eqref{Eq:FreeEnergy:London:1}, 
in the absence of phase gradients, the gauge field $\A$ is related to the 
dimensionless current $\jh$ as $\A=-\lambda_L\jh$. Thus, the dimensionless 
helicity \Eqref{Eq:Helicity:0} reads as  
$\tilde{\cal H} \equiv {\cal H} / \lambda_L =-\int \jh\cdot\B$. Here again, 
given the relations~\Eqref{Eq:London:FF:Physical} between the physical fields 
and the force-free field $\Q$, the dimensionless helicity takes the following form:
\renewcommand\arraystretch{1.25}
\Align{Eq:Helicity}{
{\cal H}&=\int\Re(\Q)\cdot\Re(\eta\Q)=\sum_{l,m}\int r^2dr{\cal H}_{lm}\,, \\
\text{where}~&{\cal H}_{lm} =\!\sum_{\mathclap{{\bs Z} = {\bs Y}, {\bs \Psi}, {\bs \Phi}} }
	w_{lm}^{\bs Z\!}\left\{
    \begin{array}{ll}
	\Re (\eta Q_{lm}^{\bs Z\!})\Re (Q_{lm}^{\bs Z\!}) 
	& \mbox{if $m$ even}\,,	\\
 	\Im (\eta Q_{lm}^{\bs Z\!})\Im (Q_{lm}^{\bs Z\!}) 
	& \mbox{if $m$ odd}\,.	
   \end{array}
\right.\nonumber
}

At small radius $r$, all the components of the force-free field \Eqref{Eq:FF:solution} 
are divergent:
\Equation{Eq:FF:diverge}{
Q_{lm}^{\bs \Phi\!} \sim r^{-(l+2)}\,,~~ 
Q_{lm}^{\bs    Y\!} \sim r^{-(l+2)}\,,~~ 
Q_{lm}^{\bs \Psi\!} \sim r^{-(l+1)}\,.
}
Therefore, all the toroflux modes, in the London limit, have an intrinsic divergence at 
the origin. For example, the divergence of the $l=1$ solution behaves as a pointlike 
magnetic dipole which, in realistic circumstances, can be regularized by the size of a 
ferromagnetic (spherical) inclusion that represents a physical dipole. The same statement 
can also be applied to the other, quadrupole ($l=2$) and higher modes. 

The conventional vortices, in the London limit, are known to have a similar divergence, 
which is resolved by introducing a cut-off at a short distance from the vortex center. 
Introducing such a cut-off is relevant because vortices have a nonzero phase winding, 
and consequently, the complex field of the condensate must vanish somewhere. However, 
these arguments cannot be applied to the case of the toroflux for the simple reason 
that the source term associated with the vortex fields is set to zero, $\vv=0$ 
[see the discussion after Eq.~\Eqref{Eq:London:j}].
Therefore, no topological arguments \emph{demand} vanishing the superconducting density, 
leading to a breakdown of the London limit that further requires the introduction of an 
ultraviolet cut-off. 

Below we consider a general case of a magnetized inclusion which naturally regularizes 
the divergence of the toroflux modes~\Eqref{Eq:FF:diverge}. 

\section{Magnetized inclusion}
\label{Sec:Inclusion}

In order to account for the divergences of the force-free field, it is instructive to 
consider the case of a magnetized (spherical) inclusion in the bulk of the 
noncentrosymmetric material. The Maxwell equations that determine the magnetic field 
inside the inclusion are 
\Equation{Eq:AM:Inclusion}{
\tCurl\vH=0	\,,\quad \tDiv\B=0\,,
~~\text{where}~~
\B=\vH+4\pi\M\,.
}
The magnetic field $\B$ and the magnetization $\M$ are decomposed onto the vector spherical 
harmonics, similarly to the force-free field $\Q$ \Eqref{Eq:VSH:decomposition}. The fields 
of the magnetized spherical inclusion are constructed following the standard textbook 
calculations (see, \eg, \cite{Jackson}, for a detailed derivation, see 
Appendix~\ref{App:Inclusion}). The general solutions are constrained by the requirement 
that the magnetic field should be a real-valued quantity, while the magnetic fields 
$\check{\B}$ and $\check{\vH}$ inside the magnetized spherical inclusion of radius $r_0$ 
satisfy the following relations:
\SubAlign{Eq:Inclusion:Solution}{
\check{H}_{lm}^{\bs Y\!}&=\check{H}_{lm}^{\bs \Psi\!}=
-\frac{4\pi l\check{M}_{lm}^{\bs Y\!}}{2l+1}
\left(\frac{r}{r_0}\right)^{l-1}\!,~\check{H}_{lm}^{\bs \Phi\!}=0,
\\
\check{B}_{lm}^{\bs Z\!}&=\check{H}_{lm}^{\bs Z\!}+4\pi\check{M}_{lm}^{\bs Z\!}
\,,~~~\text{with}~{\bs Z} = {\bs Y}, {\bs \Psi}, {\bs \Phi}\,.
}
The continuity conditions for the current and the magnetic fields at the interface 
between a magnetized inclusion inside a superconducting medium read as:
\SubAlign{Eq:BC:continuity}{
0     =&~ \J\cdot{\bs n}_{12}\,\big\vert_{r=r_0}\,,	\\
0     =&~ {\bs n}_{12}  \cdot \left( \B_2 - \B_1 \right)\,\big\vert_{r=r_0}\,, \\
\J_S  =&~ {\bs n}_{12} \times \left( {\bs H}_2 - {\bs H}_1 \right)\,\big\vert_{r=r_0}\,.
}
Here, ${\bs n}_{12}$ is the normal vector from medium 1 (the magnetized inclusion) 
to medium 2 (the parity-breaking superconductor) and $\J_S$ is the surface current density 
that is localized at the interface. The first equation in Eqs.~\Eqref{Eq:BC:continuity} 
represents the requirement of the absence of a flow of $\J$ through the interface between 
the superconductor and the magnetized inclusion \cite{Agterberg-2012}.
Using the representation \Eqref{Eq:VSH:decomposition} of the solution in the basis of the 
vector spherical harmonics, and given that ${\bs Y}_{lm}$ is the only vector harmonic that 
has a radial component, we represent the first two equations in Eqs.~\Eqref{Eq:BC:continuity} 
as 
\SubAlign{Eq:BC:continuity:1}{
\J\cdot{\bs n}_{12}\,\big\vert_{r=r_0}&=\sqrt{1-\Gamma^2}\sum_{l,m}\Im\big( 
		\eta Q_{lm}^{\bs Y\!}Y_{lm}\big)=0	\,,\\
\left( \B- \check{\B} \right)\,\big\vert_{r=r_0}&=-\sum_{l,m}\Re\big[ 
		(\eta Q_{lm}^{\bs Y\!}+ \check{B}_{lm}^{\bs Y\!})Y_{lm}\big]	=0\,.
}
Note that the intrinsic degrees of freedom of the solutions of these equations always allow 
one to reconstruct the real-valued magnetic field \Eqref{Eq:Inclusion:Solution} inside 
the inclusion. In other words, it is always possible to find the field $\check{\B}$ 
such that $\Im\check{\B}=0$. Hence, the interface conditions~\Eqref{Eq:BC:continuity:1}, 
for a given $(l,m)$ mode, boil down to
\Equation{Eq:BC:continuity:2}{
\eta Q_{lm}^{\bs Y\!}+ \check{B}_{lm}^{\bs Y\!}\,\big\vert_{r=r_0}=0 \,.
}
Finally, the use of the explicit form of the solutions for the radial functions 
\Eqref{Eq:FF:solution} and the expressions for the fields inside the spherical inclusion 
\Eqref{Eq:Inclusion:Solution}, provides us with the matching conditions that fixes the 
coefficients $c_{lm}$ as 
\Equation{Eq:BC:coefficient}{
c_{lm} =\frac{4\pi r_0 \check{M}_{lm}^{\bs Y\!}(r_0)}{l(2l+1)h_l^{(1)}(\eta r_0)} 
	\quad \text{for}~l>0\,.
}

\section{Toroflux induced by a dipole}
\label{Sec:Dipole}

Consider now the particular case of a spherical impurity of the radius $r_0$, with the 
magnetic dipole moment $\check{\M}$ directed along the axis $\hat {\bs z}$. In spherical 
coordinates, the magnetic moment of the impurity reads as follows:
\Align{Eq:Ferromagnetic}{
\check{\M} & =  M_0 {\hat {\bs z}} 
	= M_0 \left( {\bs{\hat r}}\cos\theta-{\bs{\hat\theta}}\sin\theta\right) \nonumber \\
  & =  \sqrt{\frac{4 \pi}{3}} M_0 \left( {\bs Y}_{10} + {\bs \Psi}_{10} \right)\,.
}
The continuity conditions~\Eqref{Eq:BC:coefficient} fix the only nonzero 
coefficient $c_{10}$ of the force-free field $\Q$ \Eqref{Eq:VSH:decomposition}:
\Equation{Eq:Ferromagnetic:1}{
c_{10} =\frac{ r_0 M_0 }{h_1^{(1)}(\eta r_0)}	\left(\frac{4 \pi}{3}\right)^{3/2}	\,.
}

The behavior of the Hankel functions for small arguments implies that 
\Equation{Eq:Ferromagnetic:2}{
c_{10} = i\sqrt{\frac{4\pi}{3}}\left(\frac{4\pi r_0^3}{3}\right)M_0\eta^2 
\,,\quad\text{when}~r_0\to 0	\,.
}
Thus, for a point-like dipole with the magnetic moment 
\Equation{Eq:M0d}{
M_0^{d}=\frac{4 \pi}{3}r_0^3 M_0\,,
}
the coefficient is uniquely determined as
\Equation{Eq:Dipole}{
c_{10} = i\sqrt{\frac{4\pi}{3}}\eta^2M_0^{d}	\,.
}
The quantity $M_0$ is the density of the magnetic moment in the impurity calculated per 
unit volume. The related force-free field $\Q$ corresponds to the $(l,m)=(1,0)$ harmonics:
\Align{Eq:FF:explicit}{
\Q_{10} &= -M_0^{d}\frac{\Exp{i\eta r}}{\eta r^3}
\Big[
(1-i\eta r)\big( 2\cos\theta{\bs{\hat r}} +\eta r\sin\theta{\bs{\hat \phi}} \big) 
	\nonumber	\\
&~~~~+\big(1-i\eta r(1-i\eta r)\big)\sin\theta{\bs{\hat \theta}} \Big]\,,
}
where we used the explicit form of spherical Hankel functions of the first 
kind~\Eqref{Eq:FF:solution} in order to express the solution in the closed form.
An alternative derivation via the Chandrasekhar-Kendall method is briefly outlined in 
Appendix~\ref{App:CK:approach}.
The physical fields can be reconstructed from the force-free field \Eqref{Eq:FF:explicit} 
by using the relations \Eqref{Eq:London:FF:Physical} 
\begin{widetext}
\SubAlign{Eq:explicit}{
\vH = M_0^{d} e^{-r\sqrt{1-\Gamma ^2}}&\left\lbrace
\frac{2\cos\theta }{r^3}\left[\left(\sqrt{1-\Gamma ^2}+r\right) \cos \Gamma r
						-\Gamma\sin\Gamma r\right]\,{\bs{\hat r}}
		\right.\nonumber	\\
&~~+\frac{\sin\theta }{r^3}\left[\left((1+r^2)\sqrt{1-\Gamma ^2}+r\right) \cos \Gamma r
	+\Gamma (r^2-1) \sin \Gamma  r \right]\,{\bs{\hat \theta}}	\nonumber\\
&~~\left. +\frac{\sin\theta }{r^2}	\left[\Gamma  r \cos \Gamma r
		-\left(1+r\sqrt{1-\Gamma ^2}\right) \sin \Gamma r\right]\,{\bs{\hat \phi}}	 
		\right\rbrace	\,,\\
\J =  M_0^{d}  e^{-r\sqrt{1-\Gamma ^2}}&\left\lbrace
\frac{2\cos\theta }{r^3}\left[ \Gamma r \cos\Gamma r
		-\left(1+r\sqrt{1-\Gamma ^2}\right) \sin\Gamma r\right]\,{\bs{\hat r}}
		\right.\nonumber	\\
&~~+\frac{\sin\theta }{r^3}\left[
	\Gamma r\left(1+2r\sqrt{1-\Gamma ^2}\right) \cos \Gamma r
	-\left(1+r\sqrt{1-\Gamma ^2}+r^2(1-2 \Gamma ^2)\right) \sin\Gamma r
 \right]\,{\bs{\hat \theta}}	\nonumber\\
&~~\left. -\frac{\sin\theta }{r^2}	\left[
	\Gamma  \left(1+2r\sqrt{1-\Gamma ^2}\right) \sin\Gamma r
	+\left(r(1-2\Gamma ^2)+\sqrt{1-\Gamma ^2}\right) \cos\Gamma r
	\right]\,{\bs{\hat \phi}}	 \right\rbrace\,.
}
\end{widetext}
The complex parameter $\eta$ depends on the parity-breaking parameter 
$0 \leqslant \Gamma \leqslant 1$ as $\eta=\Gamma+i\sqrt{1-\Gamma^2}$. Thus all the fields 
are exponentially localized at large distances as $\Exp{-r\sqrt{1-\Gamma^2}}$. Hence, 
the size of the toroflux, 
\Equation{Eq:L:tor}{
L_{\mathrm{tor}} = \frac{\lambda_L}{\sqrt{1 - \Gamma^2}}\,, 
}
is determined by the London penetration length $\lambda_L$ and the dimensionless 
parity-breaking coupling $\Gamma$ defined in Eq.~\Eqref{Eq:FreeEnergy:London:1}. 
In the limit of the maximal parity violation, $\Gamma \to 1$, the size of the toroflux 
diverges.

The present construction of the toroflux solutions, induced by a magnetized source in 
a noncentrosymmetric superconductor, is done in the London limit approximation, where the 
amplitude of the superconducting condensate is fixed. This implies that only the magnetic 
energy and kinetic energy of supercurrents are taken into account and thus assumes that 
there is no suppression of the order parameter. In the textbook case of conventional vortices, 
the London approximation breaks down near the vortex core, where the kinetic and magnetic 
energy densities are large enough to suppress the superconducting condensate. A similar 
situation should also be realized for the toroflux, and thus the kinetic energy density of 
supercurrents could not grow indefinitely. Instead, when the sum of the kinetic and magnetic 
energy densities becomes comparable to condensation energy density, the modulus of the order 
parameter should get suppressed $|\psi({\bs r})| < \psi_0$. This results in the formation 
of a core like structure where the growth of kinetic energy is limited by the depletion of 
the density of the order parameter.

It is instructive to estimate when the London approximation breaks down. To this end, we 
compare the energy densities $\F$, Eqs.~\eq{Eq:FreeEnergy}, of two possible configurations. 
First is $\F_1$, the energy density of the toroflux in the London approximation where 
$\psi = \psi_0$, while the second is $\F_2$, the energy of a normal state ($\psi = 0$), 
with the same magnetic field configuration. Hence the London approximation can be expected 
to break where $\F_1 > \F_2$. The energy $\F_2$ is thus 
\Equation{Eq:F2}{
\F_2 = \frac{\B^2}{8\pi} + \frac{\beta \psi_0^4}{2} = 
\frac{(M_0^d)^2 [5 + 3 \cos(2 \theta)]}{16 \pi r^6} + \frac{\beta \psi_0^4}{2}\,,
}
where the magnetic field $\B$ is that of the toroflux solution given by~\eq{Eq:FF:explicit} 
inserted into~\eq{Eq:London:FF:Physical}. The energy $\F_1$, in the London approximation,
is obtained from \eq{Eq:FreeEnergy:London:1} [again by inserting~\eq{Eq:FF:explicit} 
into~\eq{Eq:London:FF:Physical}]:
\Equation{Eq:F1}{
\F_1 = \frac{(M_0^d)^2 [5 + 3 \cos(2 \theta)]}{16 \pi r^6} 
\left( 1 + \frac{\nu^2}{\lambda_0^2}  \right)\,.
}
The requirement that $\F_1 > \F_2$ can then be cast into the following inequality:
\Equation{Eq:F1gF2}{
r^6 < \frac{(M_0^d)^2 \nu^2 [5 + 3 \cos(2 \theta)]}{8 \pi \beta \psi_0^4 \lambda_0^2}\,.
}

Now, we want to ensure that our London approximation works reasonably everywhere outside 
the magnetic inclusion. Hence the cutoff $r_c$, which corresponds to the maximal possible 
radius $r$ of the impurity where the London theory breaks down, is obtained from 
Eq.~\eq{Eq:F1gF2} with $\theta = 0$:
\Equation{Eq:London cutoff}{
r_c = \left( 4 e M_0^d \nu \xi \right)^{1/3}\,.
}
It is interesting to note that the cutoff scale in the case of a toroflux is not simply 
$\propto\xi$ as for usual vortices, but is $\propto\xi^3$ with a dependence on the size 
and magnetization of the impurity. Then our approximation will work everywhere if $r_c<r_0$, 
which results in the following condition, expressed via the parameters of the 
model~\eqref{Eq:LengthScales}:
\Equation{Eq:validity condition}{
\frac{16 \pi}{3} \nu \xi e M_0 < 1\,,
}
or, equivalently,
\Equation{Eq:validity condition:2}{
\frac{32 \pi^2 M_0 \xi \sqrt{\lambda_L^2 (1 - \Gamma^2) - \lambda_0^2}}{3 \Phi_0} < 1\,.
}
Note that this condition can always be satisfied for sufficiently small magnetic moment $M_0$.

\begin{figure*}[!htb]
\hbox to \linewidth{ \hss
\includegraphics[width=0.9\linewidth]{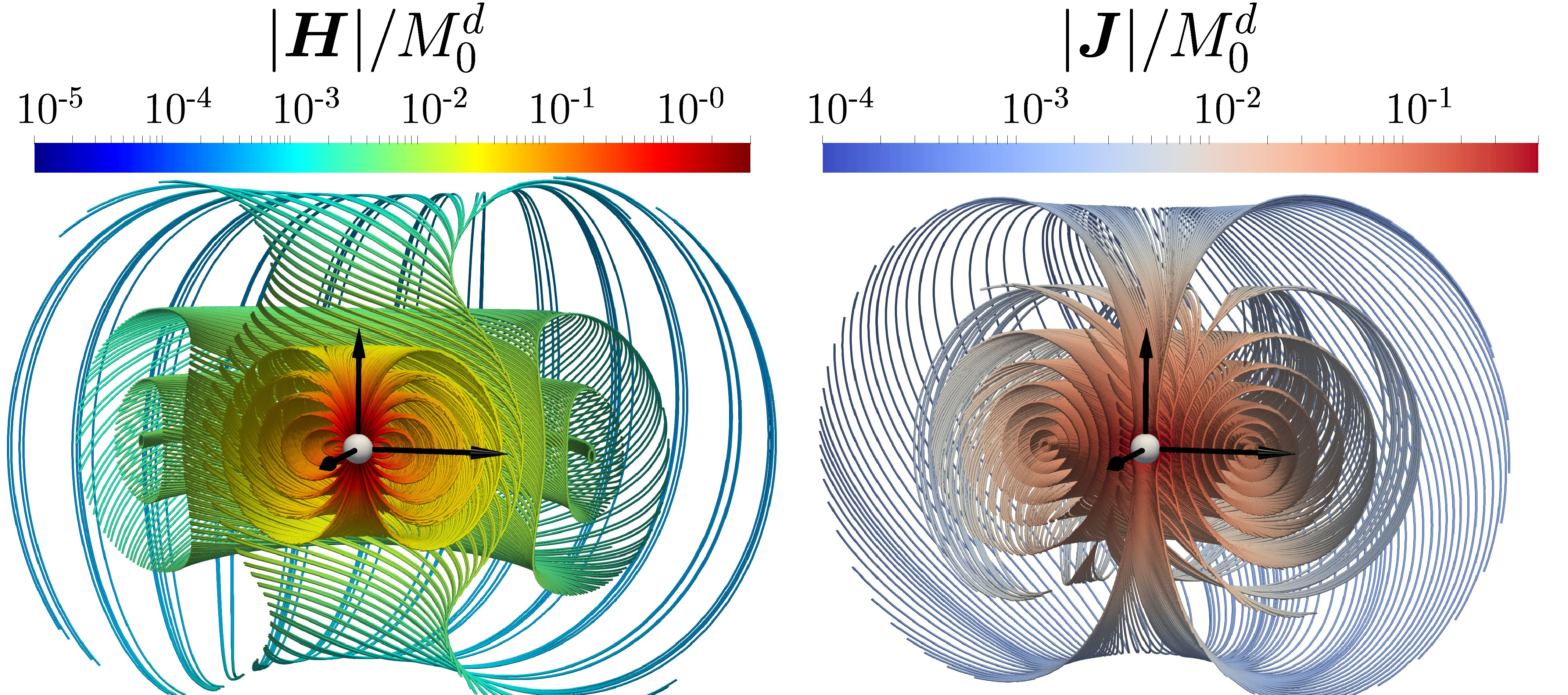}
\hss} 
\caption{ 
A toroflux solution induced by a magnetic dipole for the parity-breaking parameter 
$\Gamma=0.5$. The left panel displays the streamlines of the magnetic field $\vH$, 
while the right panel shows the streamlines of the total electric current $\J$. 
These quantities are related to each other via the Amp\`ere-Maxwell equation 
\Eqref{Eq:AM:London}. The sphere in the center shows the position of the magnetized 
inclusion (the magnetic dipole).
}
\label{Fig:Spheromak:1}
\end{figure*}

\subsection{Knotted nature of the toroflux}

The physical fields $\vH$ and $\J$ \Eqref{Eq:explicit} associated with the force-free 
field $\Q_{10}$ induced by a magnetic dipole \Eqref{Eq:FF:explicit} are displayed in 
\Figref{Fig:Spheromak:1}, for the value of the parity-breaking parameter $\Gamma=0.5$. 
This figure illustrates that a magnetic dipole impurity induces, in a noncentrosymmetric 
superconductor, the knotted lines of both the magnetic field and the electric current. 
These toroidal, axially symmetric, nested structures resemble in many aspects 
the standard Chandrasekhar-Kendall states~\cite{Chandrasekhar.Kendall-1957}. 
The alternative derivation of our solutions, presented in Appendix~\ref{App:CK:approach}, 
highlights the proximity of the toroflux and the Chandrasekhar-Kendall states. 
Note that since the magnetic lines of the toroflux are closed, the total flux 
through any cross section of the solution vanishes identically.

\begin{figure*}[p]
\hbox to \linewidth{ \hss
\includegraphics[width=0.825\linewidth]{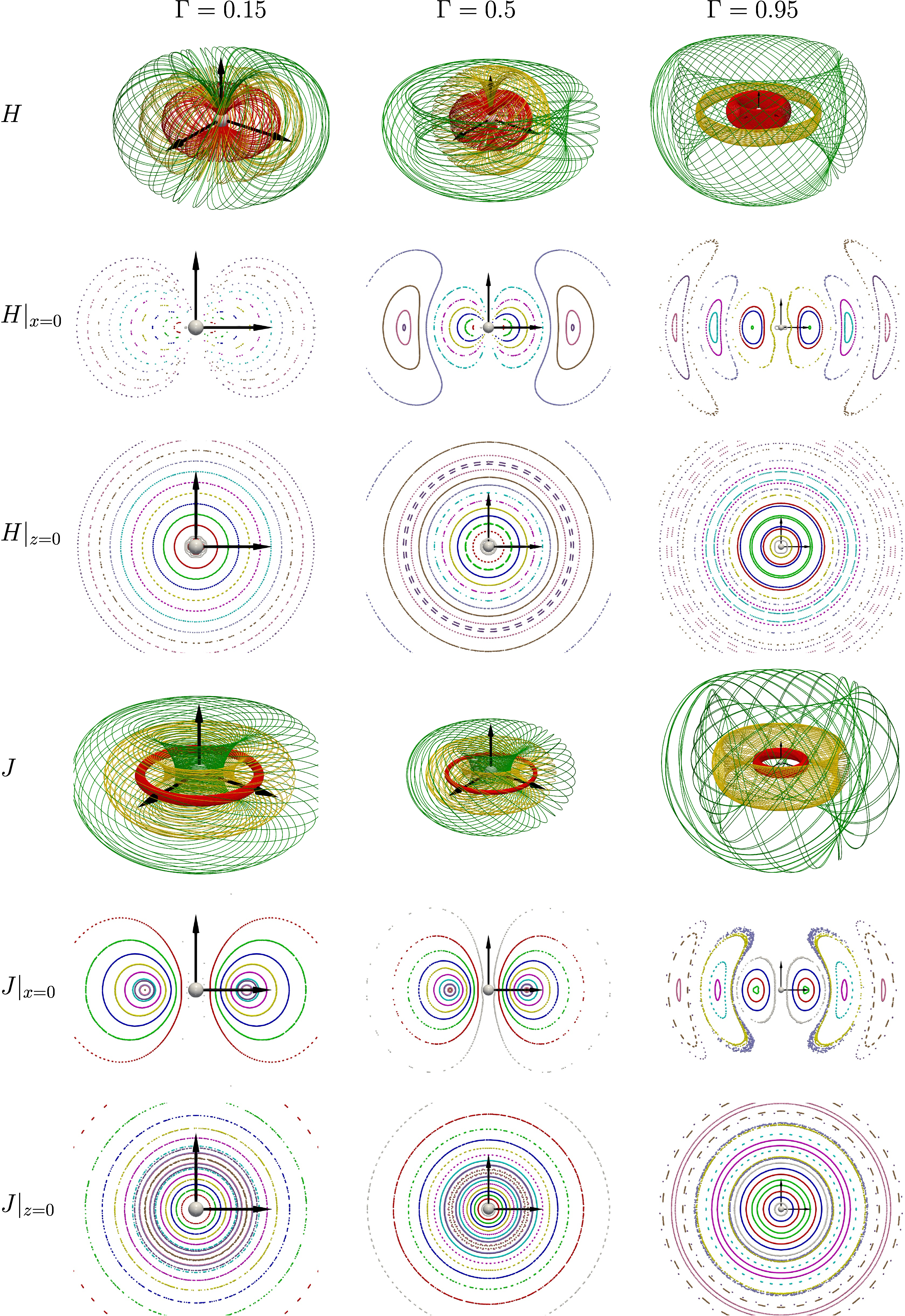}
\hss} 
\caption{ 
The structure of the streamlines of the magnetic field $\vH$ and the electric current $\J$, 
of the toroflux solution induced by a magnetic dipole, for the values of the parity-breaking 
parameter $\Gamma=0.15$, $0.5$, and $0.95$. 
The line on the top row shows the streamlines of $\vH$, and the two next rows are the 
Poincar\'e sections of the streamlines of $\vH$ on the $x=0$ and $z=0$ planes, respectively. 
Similarly, the block of the three bottom rows indicates the structure of the current $\J$. 
The central sphere depicts the spherical magnetic dipole inclusion.
%
The relative sizes of the torofluxes can be seen from the vector basis.
}
\label{Fig:Spheromak:2}
\end{figure*}

The London penetration depth determines the overall length scale of the toroflux without 
affecting the geometry of its internal structure. On the contrary, the strength of the 
noncentrosymmetricity strongly affects the overall structure of the toroflux. 
The latter feature is illustrated in Fig.~\ref{Fig:Spheromak:2}, which shows the 
streamlines of the magnetic field $\vH$ and the electric current $\J$ as well as their 
Poincar\'e sections of the torofluxes for moderate ($\Gamma=0.15$), intermediate 
($\Gamma=0.5$), and high ($\Gamma=0.95$) values of the parity-breaking parameter $\Gamma$.  

At small parity breaking ($\Gamma=0.15$), the magnetic field lines resemble that of 
a magnetic dipole. They are attached to the magnetized inclusion and slightly twisted 
around the axis of the dipole (and the chirality of twist depends on the sign of $\Gamma$). 
Accordingly, the current flows around the dipole and covers various tori. 
When the noncentrosymmetricity becomes more important ($\Gamma=0.5$), 
the toroflux features nested tori of the magnetic lines, in addition to the twisted 
structure near the dipole. This property can be seen, in particular, in the $\vH\vert_{x=0}$ 
Poincar\'e section in \Figref{Fig:Spheromak:2}. Interestingly, the chirality of the extra 
nested tori is reversed compared to the set of field lines that are attached to the dipole. 
Upon increase of the parity breaking, additional sets of nested tori appear, as can 
be seen in the $\vH\vert_{x=0}$ Poincar\'e section in \Figref{Fig:Spheromak:2} 
for $\Gamma=0.95$. The fact that the number of tori with opposite chirality increases 
as the parity breaking becomes stronger is qualitatively similar to the effect of the 
magnetic field inversion observed near vortices at large $\Gamma$ reported in 
Refs.~\cite{Garaud.Chernodub.ea-2020,Samoilenka.Babaev-2020}.

\subsection{Energy and helicity of the toroflux}

The dimensionless energy~\Eqref{Eq:Energy} of the toroflux solution \Eqref{Eq:FF:explicit} 
induced by a magnetic dipole depends on the parity-breaking parameter $\Gamma$ as 
\Align{Eq:Energy:dipole}{
\tilde{F}(\Gamma,r_0)&=\frac{2 (M_0^{d})^2}{r_0^3}\Exp{-2r_0\sqrt{1-\Gamma^2}}\Big[
(1+2r_0^2)(1-\Gamma^2)	\nonumber\\
&+\big[ 2(1-\Gamma^2)+r_0^2 \big]r_0\sqrt{1-\Gamma^2}	
\Big] \,.
}
The exponential prefactor contains the ratio of the inclusion radius $r_0$ with the 
size~\Eqref{Eq:L:tor} of the toroflux, which is the consequence of the Meissner effect.

\begin{figure}[!t]
\hbox to \linewidth{ \hss
\includegraphics[width=0.9\linewidth]{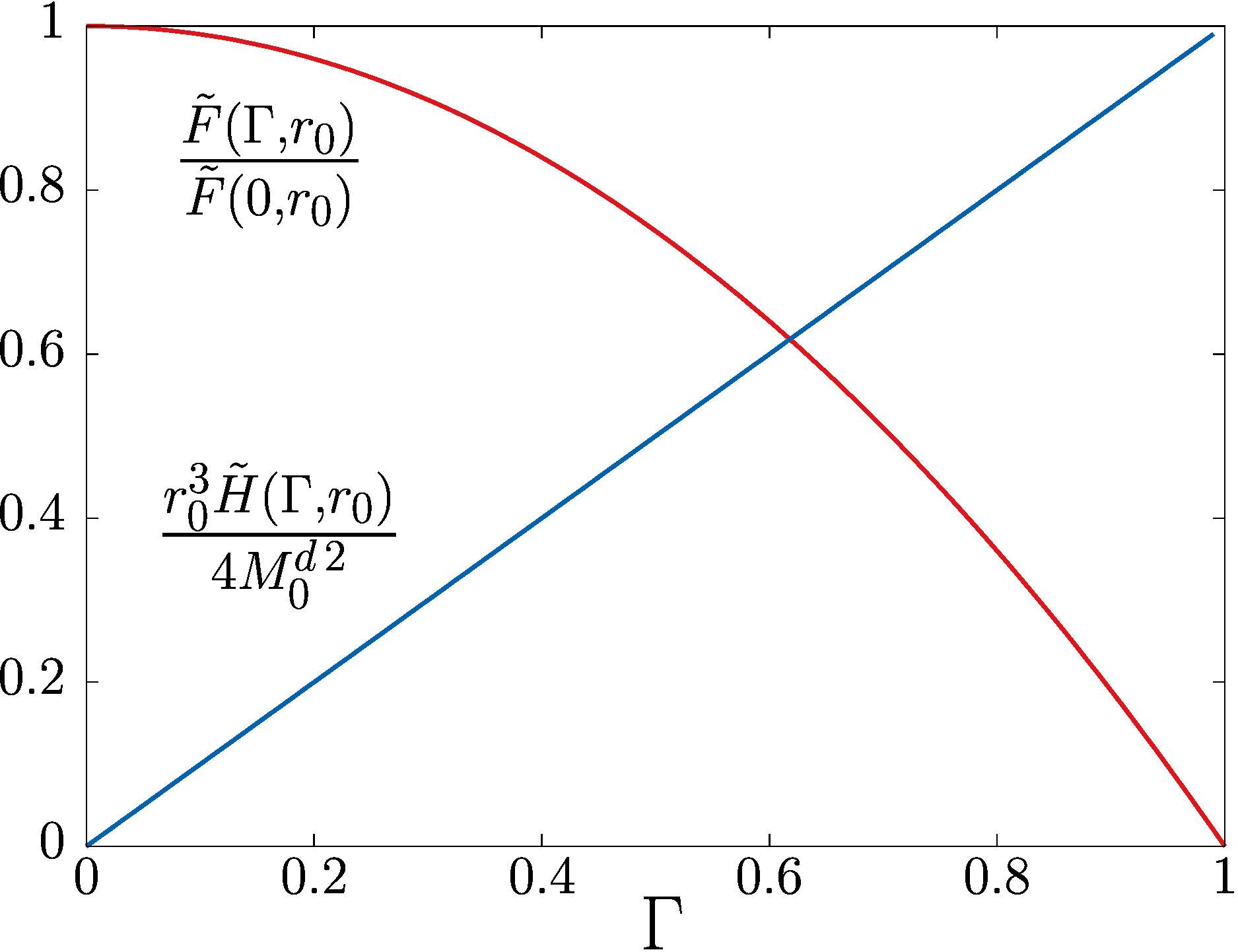}
\hss} 
\caption{
The normalized free energy $\tilde F$ and the normalized helicity $\tilde H$ of the toroflux 
as functions of the parity breaking coupling $\Gamma$ for the spherical magnetic dipole 
impurity of the radius $r_0=10^{-2}$ (in units of $\lambda_L$).
}
\label{Fig:Energy}
\end{figure}

Figure~\ref{Fig:Energy} shows the toroflux energy~\Eqref{Eq:Energy:dipole} as a 
function of the parity-breaking parameter $\Gamma$. The toroflux energy monotonically 
decreases as the parity breaking parameter $\Gamma$, and it is maximal in the 
centrosymmetric limit, $\Gamma \to 0$. This property is a consequence of the Lifshitz term 
${\bs j} \cdot {\bs B}$, which provides a negative contribution when the electric current 
and the magnetic field are (partially) aligned. When $\Gamma$ reaches the upper bound, 
$\Gamma = 1$, the toroflux energy vanishes while  its size~\Eqref{Eq:L:tor} diverges. 
Note that unlike vortices, which carry a quantized magnetic flux, the toroflux has a zero 
net flux through any plane that intersects the magnetic dipole impurity. The amplitude of 
the magnetic field is determined only by the magnetization of the dipole, hence the energy 
of our toroflux solution is not quantized.

Figure \ref{Fig:Energy} also displays the helicity~\Eqref{Eq:Helicity} of the toroflux, 
which, unlike the energy, monotonically grows with the increase of the parity-breaking 
parameter~$\Gamma$. As the nonlinear corrections are small for small (with respect to 
the London penetration depth) inclusions $r_0 \ll 1$, the helicity is almost a linear 
function of $\Gamma$. The leading contribution to the helicity at the small radius $r_0$ 
reads explicitly as follows:
\Align{Eq:Helicity:dipole}{
\frac{{\mathcal H}(\Gamma,r_0)}{2 (M_0^{d})^2}&=
\frac{2 \Gamma [1+r_0^2(\Gamma^2+1)]}{r_0^3}
-2 \left(\Gamma ^5+4 \Gamma ^3-3 \Gamma \right)  \nonumber\\
&+\frac{8 \Gamma ^5 + 12 \Gamma ^3 - 17 \Gamma }{3 \sqrt{1 - \Gamma ^2}}
+O(r_0^2)\,.
}

As previously stated, the magnetic helicity, which is associated with the topological 
properties of the magnetic field lines, serves as the measure of the linking of knotted 
lines of the magnetic field $\B$. In this respect, it is worth mentioning that the helicity 
of magnetic fields~\eq{Eq:Helicity:0} cannot be associated with an ordinary topological 
charge because this quantity is not quantized in terms of integer numbers. However, 
magnetic helicity is a topological quantity as it has a topological origin: a measure of 
the degree of linking of magnetic field lines. The non-quantization appears as a result of 
the standard definition~\eq{Eq:Helicity:0}, which scales quadratically with the magnitude 
of magnetic field (${\mathcal H} \to \alpha^2 {\mathcal H} $ if the gauge field is scaled 
by a factor of $\alpha$ as ${\bs A} \to \alpha {\bs A}$). The helicity is a very useful 
characteristic of the magnetic field in force-free environments, represented, for example, 
by ideal magnetohydrodynamics in perfectly conducting plasmas (e.g., in the solar 
corona)~\cite{Taylor-1986,Berger-1999}, which are ungapped counterparts of noncentrosymmetric 
superconductors.

Notice that the limit of a vanishing magnetic helicity, $\Gamma \to 0$, smoothly connects 
the toroflux ($\Gamma \neq 0$) with a topologically trivial screened magnetic dipole solution 
($\Gamma = 0$). Despite the energy of the solution does not vanish and does not become 
singular, the toroflux solution disappears because, at $\Gamma = 0$, the magnetic helicity 
of the configuration expectedly vanishes, and the toroflux solution becomes the usual 
dipole field screened by the superconducting condensate.

\subsection{Toroidal dipole moments}

Beyond the magnetic helicity, the toroflux solutions can be characterized by additional 
global quantities: the toroidal dipole moments. The multipole expansions are central 
to various areas of physics \cite{Thorne-1980,Dubovik.Tugushev-1990}. In particular, 
the \emph{toroidal dipole moments} have been demonstrated to play an important role in 
the electrodynamics of various condensed matter systems (for reviews, see 
\cite{Dubovik.Tugushev-1990,Spaldin.Fiebig.ea-2008,Talebi.Guo.ea-2018}).
The toroidal dipole moments of the magnetic field $\vH$ and the induction $\B$ are 
respectively defined as 
\Equation{Eq:TDM:1}{
   {\bs T}^\vH = \frac{1}{2}\int {\bs r}\times \vH \,,~~\text{and}~~~~
   {\bs T}^\B = \frac{1}{2}\int {\bs r}\times \B \,.
}

\begin{figure}[!t]
\hbox to \linewidth{ \hss
\includegraphics[width=0.9\linewidth]{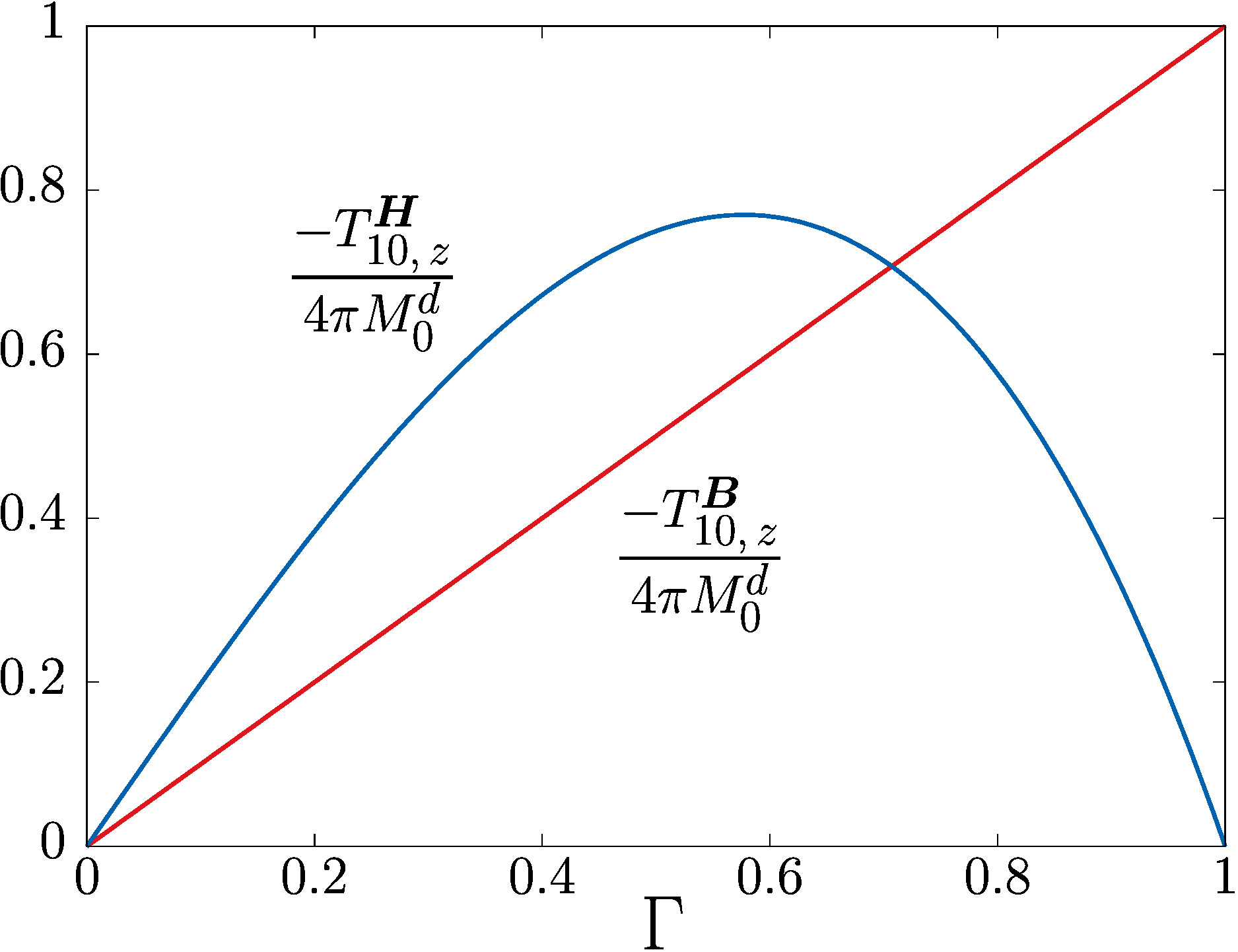}
\hss} 
\caption{
The amplitude of the toroidal dipole moments of the toroflux induced by a pointlike 
magnetic dipole, as functions of the parity-breaking coupling $\Gamma$. The toroidal 
dipole moment associated with the induction ${\bs T}_{10}^\B$ grows linearly with 
$\Gamma$, taking its maximum at the maximal value of the parity-breaking parameter, 
$\Gamma=1$. Quite surprisingly, unlike ${\bs T}_{10}^\B$, the magnetic dipole moment 
${\bs T}_{10}^\vH$ is maximal at $\Gamma=1/\sqrt{3}$, and vanishes at maximal parity 
breaking.
}
\label{Fig:Toroidal-moment}
\end{figure}

Given the relation  \Eqref{Eq:London:FF:Physical} between the physical fields
and the force-free field $\Q_{10}$ \Eqref{Eq:FF:explicit} induced by a point-like 
dipole, the toroidal dipole moments \Eqref{Eq:TDM:1} of the associated toroflux are 
(see Appendix \ref{Sec:TDM} for detailed derivation) 
\Equation{Eq:TDM:2}{
   {\bs T}_{10}^\vH = -8\pi M_0^d\Gamma(1-\Gamma^2){\hat {\bs z}} ~~\text{and}~~
   {\bs T}_{10}^\B = -4\pi M_0^d\Gamma{\hat {\bs z}} \,.
}
It follows, as illustrated in \Figref{Fig:Toroidal-moment}, that both toroidal dipole 
moments vanish in the centrosymmetric case. Interestingly, unlike ${\bs T}_{10}^\B$, 
which is a linear function of the parity-breaking parameter $\Gamma$, the toroidal 
moment ${\bs T}_{10}^\vH$ also vanishes at maximal parity breaking $\Gamma=1$.
The observation of a toroidal dipole moment of either $\B$ or $\vH$ thus 
unambiguously signals the presence of a toroflux induced by a magnetic dipole.

\subsection{Probing toroflux with muon spins}

The toroflux states could, in principle, be detected with the muon-spin spectroscopy~\cite{Schenck}. 
This method is a powerful tool to probe magnetic fields inside superconductors, allowing one 
to shed light on their superconducting properties~\cite{Kawasaki.Watanabe.ea-2013,
Singh.Hillier.ea-2014}. 

The muon-spin spectroscopy involves implanting spin-polarized positively charged muons 
(also called antimuons) in the bulk of the sample. After coming to rest inside the material, 
the spin of each muon starts to precess about the axis of the local magnetic field. 
The precession, however, does not last long, as the muon decays with a half-life of 2.2 
$\mu \mathrm{s}$. The crucial ingredient of the method is that the product of the decay, 
a positron, carries information about the direction of the muon spin at the time of the decay. 
Therefore, the measurement of the spatial distribution of the emitted positrons probes the 
local magnetic environment inside the materials. In addition, the method is susceptible to 
weak field variations, which makes it especially useful in probing the internal field structure.

The decomposition of the positron spectrum over the vector spherical harmonic modes yields 
direct information on the knottedness of the toroflux lines, which, in turn, can provide 
information about the parity-breaking parameter $\Gamma$. In our paper, we do not discuss 
the details of the experimental techniques. Instead, we assume that the spectrum gives 
information about the spherical modes of the magnetic field, in a natural extension of the 
earlier experiments~\cite{Kawasaki.Watanabe.ea-2013,Singh.Hillier.ea-2014}.

Since the experiment detects the volume-averaged flux of positrons, the positron spectrum 
is insensitive to the position of the toroflux inside the sample but is susceptible to the 
toroflux orientation. Setting the coordinate system along the axial symmetry vector of the 
toroflux, the magnetic moments of the magnetic field $\vH$ of the toroflux solution are 
\begin{equation}
    a^{\bs Z}_{lm,l'm'} = \int d^3 r\, \vH_{lm}({\bs r}) {\bs Z}_{l'm'}({\bs r})\,.
    \label{eq:a:Z}
\end{equation}
For the simplest $(l,m) = (1,0)$ solution, given explicitly in Eq.~\eq{Eq:explicit}, 
the monopole component is obviously zero, $a^{\bs Z}_{00} = 0$ (where for short we note 
$a^{\bs Z}_{l'm'} \equiv a^{\bs Z}_{10,l'm'}$). For the experimentally relevant small values 
of the radius $r_0 \ll 1$, the $(l',m')=(1,0)$ dipole-moment components take the following form:
\begin{align}
     a^{\bs Y}_{10} & =  - 4 M_0^d \sqrt{\frac{\pi}{3}} \Bigl[\Bigl(\ln r_0 {+} \gamma_E 
     {-} 1\Bigr) \cos\Gamma {+} \Gamma \sin \Gamma\Bigr] + O(r_0^2), \nonumber \\      
    a^{\bs \Psi}_{10} & =  4 M_0^d \sqrt{\frac{\pi}{3}} \Bigl[\Bigl(\ln r_0 {+} \gamma_E 
    {-} 2\Bigr) \cos\Gamma {+} \Gamma \sin \Gamma \Bigr] + O(r_0^2), \nonumber \\    
    a^{\bs \Phi}_{10} & =  8 M_0^d \sqrt{\frac{\pi}{3}} \sin \Gamma + O(r_0), \quad 
\label{eq:a:Phi:0}
\end{align} 
where $\gamma_E$ is the Euler-Mascheroni constant. In the centrosymmetric limit $\Gamma \to 0$, 
the radial, $a^{\bs Y}_{10}$, and polar, $a^{\bs \Psi}_{10}$, dipole components of the toroflux 
solution are nonvanishing. These components are characteristics of an ordinary magnetic dipole 
that possesses a trivial, unknotted magnetic field 
\footnote{Despite the apparent logarithmic divergence in the small-impurity limit, $r_0 \to 0$, 
the dipole moments $a^{\bs Y}_{10}$ and $a^{\bs \Psi}_{10}$ do not diverge since at a constant 
magnetic dipole density $M_0$, since the total dipole moment $M_0^d = M_0^d(r_0)$ diminishes 
with decreasing radius as $r_0^3$, as seen in Eq.~\eq{Eq:M0d}.
}. 
However, the presence of a nonvanishing azimuthal dipole component, 
$a^{\bs \Phi}_{10} \propto \Gamma$, which could potentially be detected in a muon-spin 
spectroscopy, can be considered as a clear sign of the presence of the knotted magnetic 
lines that are inherent to our toroflux solution. 

The nonvanishing $a^{\bs \Phi}_{10}$ component~\eq{eq:a:Phi:0} is calculated in the (laboratory) 
coordinate system in which the $z$ axis is coaligned with the magnetic dipole moment of the impurity. 
If these axes are misaligned, then, in a general coordinate system, the toroflux-sensitive 
coefficient~$a^{\bs \Phi}_{10}$ will be different from the result of Eq.~\eq{eq:a:Phi:0}. 
The difference between the system associated with the toroflux and the laboratory coordinate system 
is given by a spatial rotation with a matrix ${\hat R}$. 
Under rotation, the $(l,m)$ vector spherical harmonics transformed linearly into each other with 
the proportionality coefficients given by the Wigner $D$ matrices~\cite{stein1961addition}. 
In particular, ${\hat R} {\bs \Phi}_{lm} = \sum_{m'=-l}^l D^{l}_{mm'} {\bs \Phi}_{lm'}$. 
Substituting this equation into the definition of the magnetic moments~\eq{eq:a:Z}, we observe by 
an explicit calculation that for the $(l,m) = (1,0)$ toroflux solution~\eq{Eq:explicit}, 
the component $a^{\bs \Phi}_{10}$ is nonzero~\eq{eq:a:Phi:0}, while the other two components with 
$m = \pm 1$ vanish, $a^\Phi_{1,\pm 1} \equiv 0$. Therefore, in the rotated coordinate system, 
the ${\bs \Phi}$ magnetic dipole moment is given by 
$a^{\bs \Phi}_{10} (\vartheta) = D^1_{00}(\vartheta) a^{\bs \Phi}_{10}$, where $a^{\bs \Phi}_{10}$ 
is the magnetic moment in the local system associated with the toroflux solution and $\vartheta$ 
is the angle between the symmetry axis of the toroflux and the $z$ axis of the laboratory system. 

Taking into account the value of the corresponding Wigner matrix, $D^1_{00} = \cos\vartheta$, 
one gets 
\begin{align}
    a^{\bs \Phi}_{10} [\vartheta] =  4 M_0^d \sqrt{\frac{\pi }{3}} \Gamma \cos\vartheta\,.
    \label{eq:a:Phi:1}
\end{align}
The ${\bs \Phi}$ component of the magnetic moment~\eq{eq:a:Phi:1} has the same properties as the 
usual magnetic moment because it measures the projection of the dipole moment onto the $z$ axis. 
Its value is maximal when the $z$ axis is coaligned with the axis of the toroflux ($\vartheta = 0$), 
and it vanishes when the axess are perpendicular to each other ($\vartheta = \pi/2$). Detecting 
a nonvanishing coefficient $a^\Phi_{10}$, for example, in a muon-spin experiment, should signal 
the presence of the toroflux induced by a magnetic dipole. Since the importance of the 
$\kappa_{\pm}$ term in the Ginzburg-Landau functional strongly depends on the temperature 
\cite{Samoilenka.Babaev-2020}, it will also help differentiate the effects of 
noncentrosymmetricity from the effects caused by a random polarization of magnetic inclusions.

\section{Conclusion}

We demonstrated that noncentrosymmetric superconductors with broken inversion symmetry host 
a family of stable configurations with self-knotted magnetic field lines. These states, 
which we call \emph{toroflux}, are superconducting counterparts of the Chandrasekhar-Kendall 
states that play an important role in highly conducting, force-free plasmas relevant to 
astrophysical research and applications in nuclear 
fusion~\cite{Chandrasekhar.Kendall-1957,Taylor-1974}.
The Meissner effect forces the spatial localization of the toroflux solutions, thus making 
them different from the conventional Chandrasekhar-Kendall states.

Working in the London limit, we demonstrate that the size of the toroflux is determined by 
the London penetration length $\lambda_L$ and the dimensionless parity breaking parameter 
$0 \leqslant \Gamma \leqslant 1$, as $L_{\mathrm{tor}} = \lambda_L/\sqrt{1 - \Gamma^2}$. 
In the limit of the maximal parity violation, $\Gamma \to 1$, the size of the toroflux diverges.

The knotted nature of the toroflux states is rooted in the parity-breaking magnetoelectric 
effect that generates the supercurrent along the magnetic field lines. The supercurrent 
also produces the magnetic field, thus linking the magnetic field lines of the toroflux. 
The toroflux is characterized by a non-vanishing toroidal dipole moment of the magnetic field 
that linearly depends on the parity-breaking parameter $\Gamma$. They are also characterized 
by the helicity of the magnetic field, which vanishes in the absence of the parity breaking 
($\Gamma=0$). In this limit, the knottedness disappears, and the configuration of magnetic fields 
reduces to an ordinary dipole screened by the superconducting condensate. Since the toroflux 
is characterized by a nonvanishing magnetic helicity, in the $\Gamma \to 0$ limit, the toroflux 
solution disappears. Thus, the broken parity (with $\Gamma \neq 0)$ in a noncentrosymmetric 
superconductor plays a crucial role in the existence of the toroflux since no such configurations 
are possible in an ordinary superconductor with unbroken parity.

The torofluxes constitute an infinitely high tower of solutions labeled by orbital 
$0\leqslant l < \infty$ and magnetic $-l\leqslant m \leqslant l$ quantum numbers. 
Although the energy of any $(l,m)$-toroflux diverges at its core in the London limit, 
one could argue that toroflux energy should be finite beyond this limit (similarly to 
the energy density of conventional vortices, which is divergent in the London limit 
if a core cutoff is neglected and finite otherwise). In our paper, we focus on 
the toroflux solutions sourced by a weak magnetic inclusion. The size of the inclusion 
serves as a short-distance cutoff that regularizes the solution and makes the toroflux 
energy finite. Going beyond the London limit, for small inclusion and solutions with 
large energy density, a natural cutoff will be provided by density suppression (a corelike 
structure), with a minimal size given by the coherence length. Detailed investigation of 
this question, however, goes  beyond the scope of the current work.

We show that a finite-sized ferromagnetic inclusion with an $(l,m)$-multipole moment regularizes 
the divergences and thus induces an $(l,m)$ toroflux with finite energy. The most physically 
relevant case we discussed here in detail is the case of a magnetic dipole inclusion $(l,m)=(1,0)$. 
Note that in all generality, our solutions are regularized by any finite-size magnetized inclusion 
with a nonvanishing $(l,m)$-multipole moment.

Our findings could open up the possibility to extract new information about 
noncentrosymmetric superconductors from muon-spin rotation probes. We have demonstrated that 
the distribution of magnetic field polarization of toroflux solutions is principally different 
from a dipole field configuration of a magnetic impurity in a conventional superconductor. 
It could potentially allows us to extract the parameters $\varkappa_\pm$ from the statistics 
of the polarization of magnetic field sensed by muons.

\begin{acknowledgments}
We thank Vadim Grinenko for the discussions. A.~K. and A.~M. were supported by 
Grant No. 0657-2020-0015 of the Ministry of Science and Higher Education of Russia.
A.~M. is grateful for the kind hospitality extended to him by the Theory Group during 
his stay at the Institut Denis Poisson (Tours, France), where this work was completed. 
A.~S. and  E.~B. were supported by the Swedish Research Council Grants No. 2016-06122, 
No. 2018-03659, and No. 2022-04763.

\end{acknowledgments}

\appendix
\setcounter{section}{0}
\setcounter{equation}{0}
\renewcommand{\theequation}{\Alph{section}\arabic{equation}}


\section{Detailed derivation of a general solution via vector spherical harmonics}
\label{App:Derivation}

We derive a general solution of the force-free equation $\cL\Q=0$ using the basis of 
the spherical vector harmonics~\cite{Barrera.Estevez.ea-1985}, which provides a 
convenient separation of the radial and angular variables. The force-free field $\Q$, 
as well as the other fields, are thus decomposed as follows:
\Equation{App:Eq:VSH:decomposition}{
\Q(\x) = \sum_{l=0}^\infty \sum_{m=-l}^{+l}~\left(
\sum_{{\bs Z} = {\bs Y}, {\bs \Psi}, {\bs \Phi}} 
	Q_{lm}^{\bs Z\!}(r) \, {\bs Z}_{lm}({\hat {\bs r}}) \right) \,.
}
Here ${\bs Z}_{lm}=({\bs Y}_{lm},{\bs \Psi}_{lm},{\bs \Phi}_{lm})$ are the three 
orthogonal vector spherical harmonics, defined as \cite{Barrera.Estevez.ea-1985}
\SubAlign{App:Eq:VSH:basis}{
{\bs Y}_{lm}(   {\hat {\bs r}}) & = Y_{lm} ({\hat {\bs r}}) {\hat {\bs r}}\,, \\
{\bs \Psi}_{lm}({\hat {\bs r}}) & = r {\bs \nabla} Y_{lm} ({\hat {\bs r}})\,,\\
{\bs \Phi}_{lm}({\hat {\bs r}}) & = {\bs r} \times {\bs \nabla} Y_{lm} ({\hat {\bs r}})\,,
}
where $Y_{lm} ({\hat {\bs r}})$ are the scalar spherical harmonics which depend on
on the angular coordinates encoded in the unit vector ${\hat {\bs r}} \equiv {\bs r}/r$. 
See Appendix~\ref{App:VSH} for details on the definitions and properties of the vector 
spherical harmonics.

Given the decomposition \Eqref{App:Eq:VSH:decomposition}, the force-free vector equation 
$\cL\Q=0$ determines a system of three differential equations: 
\SubAlign{App:Eq:VSH:FF:0}{
-\frac{l(l+1)}{\eta r}Q_{lm}^{\bs \Phi\!} -\eta Q_{lm}^{\bs Y\!}  &=0	\,,\\
-\frac{1}{r}\frac{d}{dr}\left(rQ_{lm}^{\bs \Phi\!} \right)
	-\eta Q_{lm}^{\bs \Psi\!}	 &=0	\,,\\
\frac{1}{r}\frac{d}{dr}\left(rQ_{lm}^{\bs \Psi\!}\right) -\frac{1}{r}Q_{lm}^{\bs Y\!}
	-\eta Q_{lm}^{\bs \Phi\!}&= 0 \,,
}
which, combined together, yields  
\SubAlign{App:Eq:VSH:FF}{
& Q_{lm}^{\bs    Y\!} = -\frac{l(l+1)}{\eta r}Q_{lm}^{\bs \Phi\!} \,,~~ 
Q_{lm}^{\bs \Psi\!} = -\frac{1}{\eta r}\frac{d}{dr}\left(rQ_{lm}^{\bs \Phi\!}\right) 
\label{App:Eq:VSH:FF:relation}  \,,\\
&\left[ \frac{1}{r^2}\frac{d}{dr}\left(r^2 \frac{d}{dr}\right) -\frac{l(l+1)}{ r^2}
		+\eta^2\right] Q_{lm}^{\bs \Phi\!} = 0 \label{App:Eq:VSH:FF:equation} \,.
}
Equation \Eqref{App:Eq:VSH:FF:equation} on $Q_{lm}^{\bs \Phi\!}$ is the spherical 
Bessel equation whose general solution is the superposition of two spherical 
Hankel functions 
\Align{App:Eq:Q:solutions:0}{
Q_{lm}^{\bs \Phi\!} = c_{lm} h^{(1)}_l(\eta r) + d_{lm} h^{(2)}_l(\eta r).
}
Here, $h^{(1)}_l$ and $h^{(2)}_l$ are ,respectively, the Hankel functions of the first and 
second kind.

\subsection{Hankel functions}

The spherical Hankel functions are expressed via the spherical Bessel functions as 
\cite{Olver.Lozier.ea-2010} 
\Equation{App:Eq:Hankel}{
h^{(1)}_l(z)=j_l(z)+iy_l(z)\,,~~h^{(2)}_l(z)=j_l(z)-iy_l(z)\,,
}
where, in turn, the spherical Bessel functions are related to the Bessel functions 
of half-integer order:
\Equation{App:Eq:Bessel}{
j_l(z)=\sqrt{\frac{2\pi}{z}}J_{l+1/2}(z)\,,~~
y_l(z)=\sqrt{\frac{2\pi}{z}}Y_{l+1/2}(z)\,,
}
with $J_l$ and $Y_l$ being respectively the Bessel functions of the first and second kind.
Note that for a non-negative rank $l$, the spherical Hankel function of the first kind 
can be expressed in a closed form, 
\Equation{App:Eq:Hankel:closed}{
h^{(1)}_l(z)=(-i)^{l+1}\frac{\Exp{iz}}{z}\sum_{p=0}^l (-i2z)^{-p} \frac{(l+p)!}{p!(l-p)!}\,,
}
and the Hankel function of the second kind can be obtained in a similar way using the 
definition in Eq.~\Eqref{App:Eq:Hankel}.

\subsection{Asymptotics}

For large $|z|$, when $-\pi<\arg z<\pi$, the spherical Hankel functions have the 
following asymptotic behavior at the large argument~\cite{Olver.Lozier.ea-2010}:
\SubAlign{App:Eq:Hankel:asymptotics:0}{
h^{(1)}_l(z)&\sim\frac{2\pi}{z}\exp\left[i\left(z-\frac{(l+1)\pi}{2}\right)\right]\,, \\
h^{(2)}_l(z)&\sim\frac{2\pi}{z}\exp\left[-i\left(z-\frac{(l+1)\pi}{2}\right)\right]\,. 
}
By definition, $\eta=\Gamma+i\sqrt{1-\Gamma^2}$ with $\Gamma \in[0,1]$, so that
$0 < \arg (\eta r) < \pi$ and thus 
\SubAlign{App:Eq:Hankel:asymptotics}{
h^{(1)}_l(\eta r)&\sim\frac{2\pi(-i)^{l+1}}{\eta r}
		\Exp{+i\Gamma r}\Exp{-\sqrt{1-\Gamma^2}r}		\,, \\
h^{(2)}_l(\eta r)&\sim\frac{2\pi( i)^{l+1}}{\eta r} 
		\Exp{-i\Gamma r}\Exp{+\sqrt{1-\Gamma^2}r}	\,. 
}
Hence, the second spherical Hankel functions diverges at large $r$. It follows that 
for the solutions \Eqref{App:Eq:Q:solutions:0} to be bounded at infinity, we 
must have $d_{lm}=0$, and thus 
\Align{App:Eq:Q:solutions:1}{
Q_{lm}^{\bs \Phi\!} = c_{lm} h^{(1)}_l(\eta r) \,.
}
Finally, given the defining relations \Eqref{App:Eq:VSH:FF:relation}, the asymptotically 
finite components of the force-free field, associated with the different vector 
spherical harmonics are
\Align{App:Eq:FF:solution}{
& Q_{lm}^{\bs \Phi\!} = c_{lm} h_l^{(1)}(\eta r)\,,~~ 
  Q_{lm}^{\bs    Y\!} = -c_{lm}\frac{l(l+1)}{\eta r}h_l^{(1)}(\eta r) 	 \nonumber\,,\\
& Q_{lm}^{\bs \Psi\!} = -c_{lm} \left(\frac{l+1}{\eta r}h_l^{(1)}(\eta r) 
-h_{l+1}^{(1)}(\eta r)  \right) ,
}
where $c_{lm}$ is an arbitrary complex constant.

\subsection{Behaviour at small \texorpdfstring{$r$}{r}}

At small $z$, the spherical Hankel function behaves as~\cite{Olver.Lozier.ea-2010} 
\Equation{App:Eq:Hankel:origin}{
h^{(1)}_l(z)=\frac{-i2^l\Gamma(l+1/2)}{\sqrt{\pi}z^{l+1}}\,. 
}
It follows that the leading contributions of the different components of the force-free 
field at small $r$ are
\SubAlign{App:Eq:FF:origin}{
& Q_{lm}^{\bs  Y\!} = c_{lm}\frac{i2^ll(l+1)\Gamma(l+1/2)}{\sqrt{\pi}\eta^{l+2}r^{l+2}}\,,\\ 
& Q_{lm}^{\bs \Psi\!} = c_{lm}\frac{i2^ll\Gamma(l+1/2)}{\sqrt{\pi}\eta^{l+2}r^{l+2}}\,,	\\
& Q_{lm}^{\bs \Phi\!} = -c_{lm}\frac{i2^l\Gamma(l+1/2)}{\sqrt{\pi}\eta^{l+1}r^{l+1}}\,.
}
Thus, at small radius $r$, all the components of the force-free field diverge as 
\Equation{App:Eq:FF:diverge}{
Q_{lm}^{\bs \Phi\!} \sim r^{-(l+2)}\,,~~ 
Q_{lm}^{\bs    Y\!} \sim r^{-(l+2)}\,,~~ 
Q_{lm}^{\bs \Psi\!} \sim r^{-(l+1)}\,.
}
It follows that all of the toroflux modes have an intrinsic divergence at the origin, 
and therefore they require a regularization, or a cutoff, at the core of the solutions. 
We demonstrate in the next Appendix that such a regularization can consistently be done.

\section{Magnetized spherical inclusion}
\label{App:Inclusion}

\begin{figure}[!t]
\hbox to \linewidth{ \hss
\includegraphics[width=0.85\linewidth]{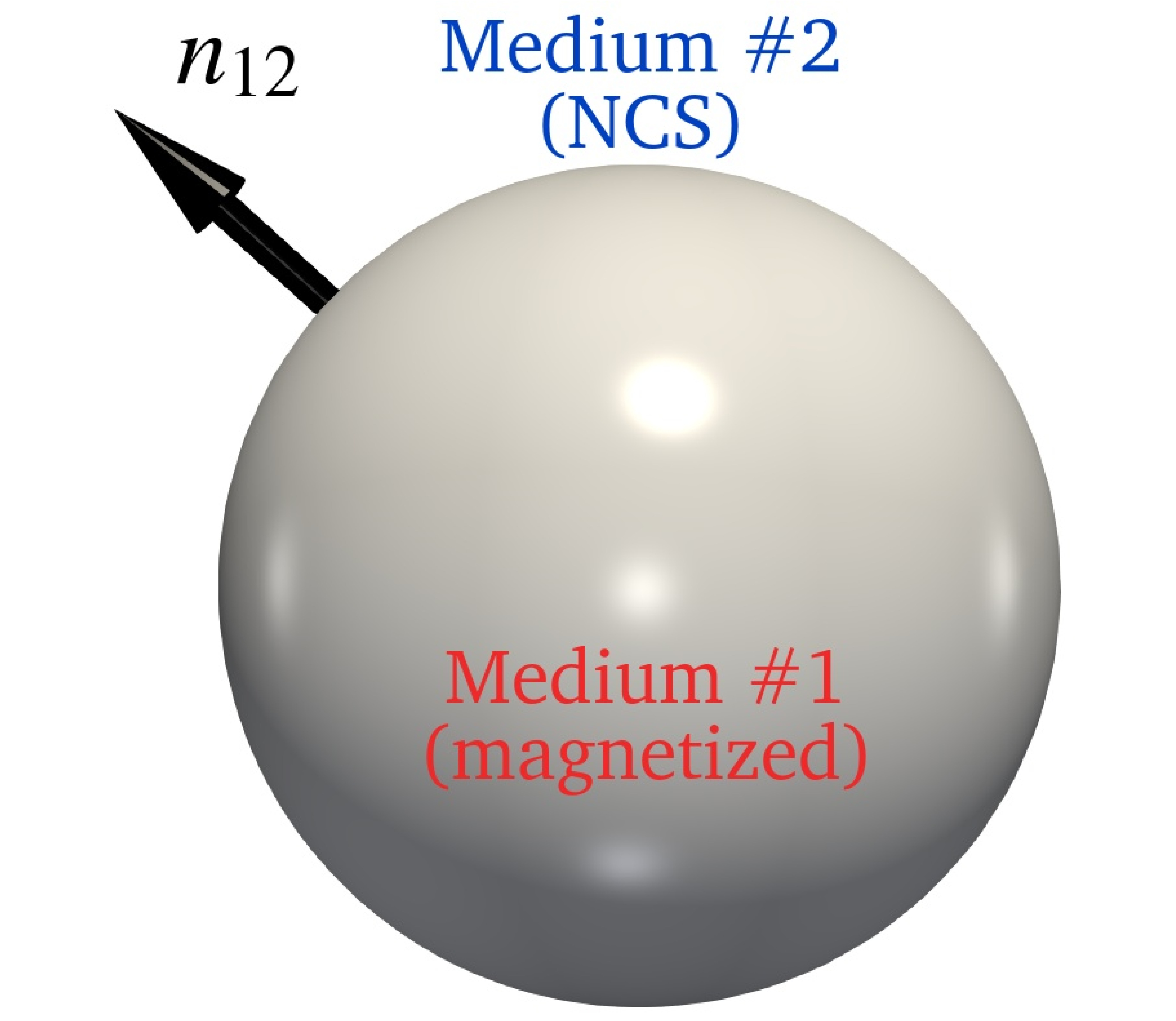}
\hss} 
\caption{ 
Schematic representation of a spherical magnetized medium (\#1) of radius $r_0$ and the 
noncentrosymmetric superconducting medium (\#2). The unit vector ${\bs n}_{12}$ is the 
normal vector at the interface between these media.
}
\label{Fig:Interface}
\end{figure}

Here, we consider the case of a magnetized inclusion in the bulk of the noncentrosymmetric 
medium. For simplicity, we study a spherical inclusion of radius $r_0$, as it is 
schematically illustrated in \Figref{Fig:Interface}. Inside a magnetized medium, 
the constituent magnetostatics equations are 
\Equation{App:Eq:AM:Inclusion}{
\tCurl\vH=0	\,,~ \tDiv\B=0 
~~\text{where}~~
\B=\vH+4\pi\M\,.
}
The fields of the magnetized spherical inclusion are constructed following the 
standard textbook calculations (see, \eg, \cite{Jackson}). To this end, we introduce 
the magnetic scalar potential $\omega_M$, which describes the magnetic field 
$\vH = -\tGrad\omega_M$, and decompose the magnetic potential over the (scalar) 
spherical harmonics. It follows that 
\SubAlign{App:Eq:H:potential}{
\vH &= -\sum_{l=0}^\infty \sum_{m=-l}^{+l}\frac{d\,\omega_{lm}}{dr}{\bs Y}_{lm}
+\frac{\omega_{lm}}{r}{\bs \Psi}_{lm} \,,\\
\text{where}~&
\left[\frac{d^2}{dr^2} +2r\frac{d}{dr}
-l(l+1)
\right]\omega_{lm}=0	\,,
}
where the second equation follows from the relation $\tDiv\vH=0$ which reflects the 
closeness of the lines of the field $\vH$. This determines the magnetic potential 
associated with the magnetized inclusion:
\Equation{App:Eq:H:potential:1}{
\omega_{lm}=\left\{
    \begin{array}{ll}
	\check{c}_{lm}r^l
	& \mbox{if $r<r_0$}	\\
 	\check{d}_{lm}r^{-(l+1)}
	& \mbox{if $r>r_0$}	\,.
   \end{array}
\right.
}
Hereafter, the symbol $\check{}$ marks the quantities inside the spherical inclusion.

The continuity of the magnetic potential at the boundary $r=r_0$ implies that 
$\check{d}_{lm}=\check{c}_{lm}r_0^{2l+1}$. The normal derivative of 
the magnetic potential is discontinuous across the interface. 
Therefore, the relation $\Div(\vH+4\pi\M)=0$ implies that, 
at $r=r_0$,  
\Equation{App:Eq:H:potential:continuity}{
\frac{d\,\omega_{lm}^{\text{in}}}{dr}-\frac{d\,\omega_{lm}^{\text{out}}}{dr}
=4\pi \check{M}_{lm}^{\bs Y\!}	~~\Rightarrow~~
\check{c}_{lm}=\frac{4\pi \check{M}_{lm}^{\bs Y\!}}{(2l+1)r_0^{l-1}}\,.
}
Here, $\check{M}_{lm}^{\bs Y\!}$ is the ${\bs Y}$ component of the magnetization 
$\check{\M}$, decomposed over the vector spherical harmonics analogously to 
\Eqref{Eq:VSH:decomposition}. Hence, the magnetic fields inside the magnetized spherical 
inclusion are 
\SubAlign{App:Eq:H:Inclusion}{
\check{H}_{lm}^{\bs Y\!}&=\check{H}_{lm}^{\bs \Psi\!}=
    -\frac{4\pi l\check{M}_{lm}^{\bs Y\!}}{2l+1}
\left(\frac{r}{r_0}\right)^{l-1}\,,~\check{H}_{lm}^{\bs \Phi\!}=0\,,
\\
\check{B}_{lm}^{\bs Z\!}&=\check{H}_{lm}^{\bs Z\!}+4\pi\check{M}_{lm}^{\bs Z\!}
\,,~~~\text{with}~{\bs Z} = {\bs Y}, {\bs \Psi}, {\bs \Phi}\,.
}
As detailed below, the interface boundary conditions between the magnetized inclusion and the 
superconductor allow one to relate the values of the parameters $c_{lm}$ between both media.

\subsection{Matching conditions at the interface}

The continuity conditions for the current and the magnetic fields at the interface 
between a magnetized inclusion inside a superconducting medium read as:
\SubAlign{App:Eq:BC:continuity}{
0     =&~ \J\cdot{\bs n}_{12}\,\big\vert_{r=r_0}\,,	\\
0     =&~ {\bs n}_{12}  \cdot \left( \B_2 - \B_1 \right)\,\big\vert_{r=r_0}\,, \\
\J_S  =&~ {\bs n}_{12} \times \left( {\bs H}_2 - {\bs H}_1 \right)\,\big\vert_{r=r_0}\,.
}
Here, ${\bs n}_{12}$ is the normal vector from medium 1 (the magnetized inclusion) 
to the medium 2 (the parity-odd superconductor) and $\J_S$ is the surface current 
density that is localized at the interface, Fig.~\ref{Fig:Interface}. 

The first equation in Eqs.~\eq{App:Eq:BC:continuity} states that the superconducting current 
$\J$ does not enter the non-superconducting magnetized inclusion \cite{Agterberg-2012}. 
Consequently, the normal component of $\J$ vanishes at the interface between these media. 
Given the decomposition \Eqref{App:Eq:VSH:decomposition} over the vector spherical 
harmonics, and since ${\bs Y}_{lm}$ is the only vector harmonic that has a radial 
component, the first two relations in Eqs.~\eq{App:Eq:BC:continuity} reduce to 
\SubAlign{App:Eq:BC:continuity:1}{
\J\cdot{\bs n}_{12}\,\big\vert_{r=r_0}&=\sqrt{1-\Gamma^2}\sum_{l,m}\Im\big( 
		\eta Q_{lm}^{\bs Y\!}Y_{lm}\big)=0	\,,\\
\left( \B- \check{\B} \right)\,\big\vert_{r=r_0}&=-\sum_{l,m}\Re\big[ 
		(\eta Q_{lm}^{\bs Y\!}+ \check{B}_{lm}^{\bs Y\!})Y_{lm}\big]	=0\,.
}
Note that there is always the freedom to construct the magnetic field $\check{\B}$ inside 
the spherical inclusion, so that it is real. It is thus always possible to construct 
$\check{\B}$ such that $\Im\check{\B}=0$. Hence, the conditions 
\Eqref{App:Eq:BC:continuity:1}
at the interface, for a given mode $(l,m)$, boil down to
\Equation{App:Eq:BC:continuity:2}{
\eta Q_{lm}^{\bs Y\!}+ \check{B}_{lm}^{\bs Y\!}\,\big\vert_{r=r_0}=0 \,.
}
Finally, we use the explicit form of the radial modes~\Eqref{App:Eq:FF:solution} 
to fix all the coefficients $c_{lm}$ of the solution:
\Equation{App:Eq:BC:continuity:3}{
c_{lm} =\frac{r_0\check{B}_{lm}^{\bs Y\!}(r_0)}{l(l+1)h_l^{(1)}(\eta r_0)} 
	~~~\text{for}~l>0\,.
}
Now, given the $(l,m)$ magnetization modes of the magnetized spherical inclusion 
\Eqref{App:Eq:H:Inclusion}, the arbitrary coefficient $c_{lm}$ reads as
\Equation{App:Eq:BC:continuity:4}{
c_{lm} =\frac{4\pi r_0 \check{M}_{lm}^{\bs Y\!}(r_0)}{l(2l+1)h_l^{(1)}(\eta r_0)} 
	~~~\text{for}~l>0\,.
}
Thus, we obtain the most general solution for a spherical inclusion with arbitrary 
magnetization.

\subsection{Ferromagnetic inclusion and magnetic dipole}

Consider now the particular case of a spherical inclusion of radius $r_0$, which possesses 
a magnetic dipole moment $\check{\M}$ directed along the axis $\hat {\bs z}$, and with all 
higher-order modes vanishing. In spherical coordinates, the magnetic moment reads as 
\SubAlign{App:Eq:Ferromagnetic}{
\check{\M} & =  M_0 {\hat {\bs z}} 
	= M_0 \left( {\bs{\hat r}}\cos\theta-{\bs{\hat\theta}}\sin\theta\right) \nonumber \\
  & =  \sqrt{\frac{4 \pi}{3}} M_0 \left( {\bs Y}_{10} + {\bs \Psi}_{10} \right)\,.
}
Thus, the magnetic fields \Eqref{App:Eq:H:Inclusion} inside the inclusion are
\SubAlign{App:Eq:Ferromagnetic:Inclusion}{
\check{H}_{10}^{\bs Y\!}&=\check{H}_{10}^{\bs \Psi\!}=
	-\left(\frac{4 \pi}{3}\right)^{3/2}M_0
\,,~\check{H}_{10}^{\bs \Phi\!}=0	\,,
\\
\check{B}_{10}^{\bs Y\!}&=\check{B}_{10}^{\bs \Psi\!}=
\,2\left(\frac{4 \pi}{3}\right)^{3/2}M_0
\,,~\check{B}_{10}^{\bs \Phi\!}=0 \,.
}
Finally, given the continuity conditions, the free coefficient $c_{10}$
\Eqref{App:Eq:BC:continuity:4} in this case becomes 
\Equation{App:Eq:Ferromagnetic:1}{
c_{10} =\frac{ r_0 M_0 }{h_1^{(1)}(\eta r_0)}	\left(\frac{4 \pi}{3}\right)^{3/2}	\,.
}

The behavior of the Hankel functions for small arguments \Eqref{App:Eq:Hankel:origin}, 
implies that 
\Equation{App:Eq:Ferromagnetic:2}{
c_{10} = i\sqrt{\frac{4\pi}{3}}\left(\frac{4\pi r_0^3}{3}\right)M_0\eta^2 
\,,~\text{when}~r_0\to 0	\,.
}
Thus, for a pointlike dipole with the magnetic moment $M_0^{d}=\frac{4 \pi}{3}r_0^3M_0$, 
the coefficient is uniquely determined as
\Equation{App:Eq:Dipole}{
c_{10} = i\sqrt{\frac{4\pi}{3}}\eta^2M_0^{d}	\,.
}

Now, given the coefficient \Eqref{App:Eq:Dipole}, the magnetic field $\vH$ and the current 
$\J$ induced by a magnetic pointlike dipole can be reconstructed from the force-free field 
$\Q$ according to \Eqref{Eq:London:FF:Physical}. The components of the force-free field $\Q$ 
corresponding to a given sector of the vector spherical harmonics are defined in terms of 
the spherical Hankel functions of the first kind \Eqref{Eq:FF:solution}. These functions 
can further be expressed in a closed form using the relation \Eqref{App:Eq:Hankel:closed}. 
Finally, the vector harmonics of a dipolar source possess the single mode $(l,m)=(1,0)$, 
which has the simple form~\Eqref{App:Eq:VSH:1}. Thus, in terms of elementary functions, 
the force-free field $\Q_{10}$ induced by a magnetic dipole reads as  
\Align{App:Eq:FF:explicit}{
\Q_{10} &= -M_0^{d}\frac{\Exp{i\eta r}}{\eta r^3}
\Big[
(1-i\eta r)\big( 2\cos\theta{\bs{\hat r}} +\eta r\sin\theta{\bs{\hat \phi}} \big) 
	\nonumber	\\
&~~~~+\big[1-i\eta r(1-i\eta r)\big]\sin\theta{\bs{\hat \theta}} \Big]\,.
}

\subsection{Toroidal dipole moments}
\label{Sec:TDM}

The multipole expansions are central in many areas of physics \cite{Thorne-1980,
Dubovik.Tugushev-1990}. A particularly interesting kind of multipole moments 
are the \emph{toroidal moments}, which play an important role in the electrodynamics 
of various condensed matter systems (for reviews, see 
\cite{Dubovik.Tugushev-1990,Spaldin.Fiebig.ea-2008,Talebi.Guo.ea-2018}).
The toroidal dipole moments of the magnetic field $\vH$ and the induction $\B$ are 
respectively defined as 
\Equation{App:Eq:TDM:1}{
   {\bs T}^\vH = \frac{1}{2}\int {\bs r}\times \vH ~~~\text{and}~~~~
   {\bs T}^\B = \frac{1}{2}\int {\bs r}\times \B \,.
}
Given the relation between the physical fields \Eqref{Eq:London:FF:Physical}
and the force-free field $\Q_{10}$ \Eqref{App:Eq:FF:explicit} induced by a pointlike 
dipole, the toroidal dipole moments \Eqref{Eq:TDM:2} of the associated toroflux 
thus read as
\Align{App:Eq:TDM:2}{
   {\bs T}_{10}^\vH &= -\frac{1}{2}\Re\left[\eta
   \sum_{{\bs Z} = {\bs Y}, {\bs \Psi}, {\bs \Phi}} 
    \int Q^{\bs Z}_{10}\,({\bs r}\times{\bs Z}_{10}) \right]   \,,\\
   {\bs T}_{10}^\B  &= \frac{\sqrt{1-\Gamma^2}}{2}\Im\left[
   \sum_{{\bs Z} = {\bs Y}, {\bs \Psi}, {\bs \Phi}} 
    \int Q^{\bs Z}_{10}\,({\bs r}\times{\bs Z}_{10}) \right] \,.\nonumber
}
It is thus necessary to evaluate the volume integral
\Equation{App:Eq:TDM:QIntegral:1}{
\int Q^{\bs Z}_{10}\,({\bs r}\times{\bs Z}_{10})=\int r^3 Q^{\bs Z}_{10} \,dr
        \int  {\bs{\hat r}}\times {\bs Z}_{10}({\hat {\bs r}}) \,d\Omega\,.
}
To this end, note the property of the vector spherical harmonics 
\SubAlign{App:Eq:Zlm:vector_prod}{
{\bs{\hat r}}\times {\bs Y}_{lm}({\hat {\bs r}})    & = 0 \,,  \\
{\bs{\hat r}}\times {\bs \Psi}_{lm}({\hat {\bs r}}) &= {\bs \Phi}_{lm}({\hat {\bs r}})  \,,\\
{\bs{\hat r}}\times {\bs \Phi}_{lm}({\hat {\bs r}}) & = -{\bs \Psi}_{lm}({\hat {\bs r}}) \,,
}
and that the only non-vanishing integral over the sphere here is 
$\int  {\bs \Psi}_{10} \,d\Omega=2\sqrt{\frac{4\pi}{3}}  {\bs{\hat z}}$
(see  details in Appendix \ref{App:Sec:VSH:itegrals}).
Moreover the radial integral also gives a simple relation 
\Equation{App:Eq:TDM:HankelIntegral}{
    \int r^3 h^{(1)}(\eta r) \,dr = \frac{3i}{\eta^4}\,.
}
Thus, given the coefficient $c_{10}$ \Eqref{App:Eq:Dipole} associated with the pointlike 
magnetic dipole, the volume integral \Eqref{App:Eq:TDM:QIntegral:1} becomes
\Equation{App:Eq:TDM:QIntegral:2}{
\sum_{{\bs Z} = {\bs Y}, {\bs \Psi}, {\bs \Phi}}
    \int Q^{\bs Z}_{10}\,({\bs r}\times{\bs Z}_{10})= 
        \big(8\pi M_0^d\eta^{*\,2} \big){\bs{\hat z}}\,.
}
Finally, using this relation together with the definition of 
toroidal dipole moments \Eqref{App:Eq:TDM:2} yields the relations 
\Equation{App:Eq:TDM:3}{
   {\bs T}_{10}^\vH = -8\pi M_0^d\Gamma(1-\Gamma^2){\hat {\bs z}} ~~\text{and}~~
   {\bs T}_{10}^\B = -4\pi M_0^d\Gamma{\hat {\bs z}} \,.
}

\section{Outline of the Chandrasekhar-Kendall approach for a dipole source}
\label{App:CK:approach}

The above section provides the explicit forms of the toroflux solutions induced 
by a dipole. These can alternatively be derived via the Chandrasekhar-Kendall method 
\cite{Chandrasekhar.Kendall-1957}. Consider a case when we want to find the magnetic field 
$\B$ that satisfies the following equation:
\Equation{App:Eq:B}{
\cL \cL^* \B = c {\bs \nabla} \times \left( {\bs \nabla} \times \M^d \right) 
             + d {\bs \nabla} \times \M^d \,,
}
where $c$ and $d$ are some real parameters and $\M^d$ is a field that corresponds to 
an external field induced by a magnetic moment.
Then, the magnetic field $\B$ can be solved in terms of the following functions:
\SubAlign{App:Eq:solution}{
\B = -\Re\left( \eta \Q \right) \,, \\
\Q = {\bs \nabla} \times {\bs u} + {\bs \nabla} \times \left( {\bs \nabla} \times {\bs u} \right) 
/ \eta\,, \label{App:Eq:solution:Q_eq} \\
\Delta {\bs u} + \eta^2 {\bs u} = b \M^d\,,\ \  
b = - i (d + c \eta) / \Im\eta\,. 
\label{App:Eq:solution:u_eq}
}
where ${\bs u}$ is found from solving the inhomogeneous vector Helmholtz equation 
\Eqref{App:Eq:solution:u_eq}.
The set of equations~\Eqref{App:Eq:solution} can be verified by showing that
\Equation{App:Eq:not_force_free}{
\cL \Q = - b {\bs \nabla} \times \M^d / \eta\,,
}
which subsequently implies Eq.~\Eqref{App:Eq:B}.

In the simplest case of pointlike dipole source 
$\M^d = M_0^{d} {\hat {\bs z}} \delta({\bs r})$, the explicit solution of the Helmholtz 
equation~\Eqref{App:Eq:solution:u_eq} is 
\Equation{App:Eq:u_solution}{
{\bs u} = - M_0^{d} {\hat {\bs z}} b \frac{e^{i \eta r}}{4 \pi r}\,.
}

Note, that values of $c$ and $b$ depend on boundary conditions that are used 
between magnetized and superconducting mediums. The boundary conditions 
\Eqref{App:Eq:BC:continuity:2} are given by 
\Equation{App:Eq:BC:QB}{
\left( \eta \Q + \check{\B} \right) \cdot {\hat {\bs r}} \big\vert_{r=r_0} = 0 \,.
}
For $\check{\B}(r_0) = 2 M_0^{d} {\hat {\bs z}} / r_0^3$ this results in  the constants 
being $c = 4 \pi$ and $d = - 4 \pi \Re \eta$ (and thus $b = 4 \pi$).

Now, given the values of $c$ and $d$ that satisfy the appropriate boundary condition, 
inserting the solution \Eqref{App:Eq:u_solution} of the Helmholtz equation into the 
constituting equation \Eqref{App:Eq:solution:Q_eq} yields the very same expression for 
the force-free field $\Q$ as the explicit form  of Eq.~\Eqref{App:Eq:FF:explicit}.

\section{Spherical harmonics}
\label{App:VSH}

The scalar spherical harmonics are defined as \cite{Olver.Lozier.ea-2010} 
\Equation{App:Eq:Ylm}{
Y_{lm} ({\hat {\bs r}}) 
=  (-1)^m \sqrt{\frac{2 l + 1}{4 \pi} \frac{(l - m)!}{(l + m)!}} 
	P_{l}^m (\cos \theta) e^{i m \phi}\,,
}
where $P_{l}^m$ are the associated Legendre polynomials. The spherical harmonics depend on 
the polar $\theta$ and azimuthal $\phi$ angles expressed collectively via the unit vector 
${\hat {\bs r}} \equiv {\bs r}/r$. The spherical harmonics satisfy the orthonormality 
condition:
\Equation{App:Eq:Ylm:orthogonality}{
\int d\Omega\, Y_{lm} ({\hat {\bs r}}) Y^*_{l'm'} ({\hat {\bs r}}) = 
		\delta_{ll'} \delta_{mm'}\,,
}
where $Y^*_{lm} \equiv (-1)^m Y_{l,-m}$ and $d \Omega = \sin\theta d\theta d \phi$ is the 
solid-angle element.

Adopting the parametrization of Ref.~\cite{Barrera.Estevez.ea-1985}, the vector spherical 
harmonics are defined via their scalar counterpart in Eq.~\eq{App:Eq:VSH:basis}.
In a given $(l,m)$ sector, the vector spherical harmonics are locally orthogonal to each 
other at every point of the unit sphere:
\SubAlign{App:Eq:Zlm:orthogonality}{
{\bs Y}_{lm}({\hat {\bs r}}) \cdot {\bs \Psi}_{lm}({\hat {\bs r}}) & =  0,  \\
{\bs Y}_{lm}({\hat {\bs r}}) \cdot {\bs \Phi}_{lm}({\hat {\bs r}}) & =  0,  \\
{\bs \Psi}_{lm}({\hat {\bs r}}) \cdot  {\bs \Phi}_{lm}({\hat {\bs r}}) & =  0.  
}
They also satisfy the normalization and orthogonality relations:
\SubAlign{App:Eq:Zlm:orthonormality}{
\int d \Omega \, {\bs Y}_{lm}({\hat {\bs r}}) 
	\cdot {\bs Y}^*_{l'm'}({\hat {\bs r}}) 	  & = \delta_{ll'} \delta_{mm'}\,, \\
\int d \Omega \, {\bs \Phi}_{lm}({\hat {\bs r}}) 
	\cdot {\bs \Phi}^*_{l'm'}({\hat {\bs r}}) & = l(l+1)\delta_{ll'}\delta_{mm'}\,, \\
\int d \Omega \, {\bs \Psi}_{lm}({\hat {\bs r}}) 
	\cdot {\bs \Psi}^*_{l'm'}({\hat {\bs r}}) & = l(l+1)\delta_{ll'} \delta_{mm'}\,, \\
\int d \Omega \, {\bs Y}_{lm}({\hat {\bs r}}) 
	\cdot {\bs \Psi}^*_{l'm'}({\hat {\bs r}}) & = 0	\,,\\
\int d \Omega \, {\bs Y}_{lm}({\hat {\bs r}}) 
	\cdot {\bs \Phi}^*_{l'm'}({\hat {\bs r}}) & = 0	\,,\\
\int d \Omega \, {\bs \Phi}_{lm}({\hat {\bs r}}) 
	\cdot {\bs \Psi}^*_{l'm'}({\hat {\bs r}}) & = 0	\,,
}
where 
\Equation{App:Eq:Zlm:conjugate}{
{\bs Z}^*_{lm} \equiv (-1)^m {\bs Z}_{l,-m}\,,
}
for ${\bs Z}_{lm} = ({\bs Y}_{lm}, {\bs \Psi}_{lm},{\bs \Phi}_{lm})$.

\subsection{Differential operators on vector spherical harmonics}

The divergence of the vector spherical harmonics is:
\SubAlign{App:Eq:VSH:Div}{
& \Div\Big(f(r){\bs    Y}_{lm}\Big) = \left(\frac{df}{dr}+\frac{2}{r}f\right) Y_{lm}, \\
& \Div\Big(f(r){\bs \Psi}_{lm}\Big) = - \frac{l(l+1)}{r} f Y_{lm}, \\
& \Div\Big(f(r){\bs \Phi}_{lm}\Big) = 0\,. 
}
Similarly, the curl of the vector spherical harmonics gives
\SubAlign{App:Eq:VSH:Curl}{
\Curl\Big(f(r){\bs   Y}_{lm}\Big)&=-\frac{1}{r} f {\bs \Phi}_{lm}, \\
\Curl\Big(f(r){\bs\Psi}_{lm}\Big)&=\left(\frac{df}{dr}+\frac{1}{r}f\right) 
			{\bs \Phi}_{lm} \\
\Curl\Big(f(r){\bs\Phi}_{lm}\Big)&=-\frac{l(l+1)}{r}f{\bs Y}_{lm} \nonumber \\
			&\quad-\left(\frac{df}{dr}+\frac{1}{r}f\right){\bs\Psi}_{lm}\,.
}
These relations allow one to express the divergence,
\Equation{App:Eq:generic:div}{
\Div{\bs G} = \sum_{l=0}^\infty \sum_{m=-l}^{l} \left(
	 \frac{1}{r^2}\frac{d}{dr}\left( r^2G^{\bs Y\!}_{lm} \right)
		- \frac{l(l+1)}{r} G^{\bs\Psi\!}_{lm} \right) Y_{lm},
}
and the curl,
\Align{App:Eq:generic:curl}{
{\bs \nabla}\times {\bs G}  = &\sum_{l=0}^\infty \sum_{m=-l}^{l} \Bigg[ 	
	- \frac{l(l+1)}{r} G^{\bs\Phi\!}_{lm} {\bs Y}_{lm}  \nonumber \\
	&- \left(\frac{d G^{\bs\Phi\!}_{lm}}{dr} 
			+ \frac{1}{r} G^{\bs\Phi\!}_{lm} \right) {\bs \Psi}_{lm} 
	\nonumber\\
	&+ \left( - \frac{1}{r} G^{\bs Y\!}_{lm} + \frac{d G^{\bs\Psi\!}_{lm}}{dr} 
		+ \frac{1}{r} G^{\bs\Psi\!}_{lm} \right) {\bs \Phi}_{lm} \Bigg],
}
of a generic vector:
\Equation{App:Eq:generic:expansion}{
{\bs G}({\bs r}) = \sum_{l=0}^{\infty} \sum_{m=-l}^{l} \left(
\sum_{{\bs Z} = {\bs Y}, {\bs \Psi}, {\bs \Phi}} 
	G_{l}^{\bs Z\!}(r) \, {\bs Z}_{lm}({\hat {\bs r}}) \right).
}

\subsection{First vector spherical harmonics}

It is useful to consider a first few spherical harmonics explicitly. Note that 
the harmonics with negative indices $m$ can be obtained from the equations below, 
using the conjugacy relation~\Eqref{App:Eq:Zlm:conjugate}.

The lowest $l=0$ harmonics is a trivial hedgehog as it possesses a radial component only:
\Equation{App:Eq:VSH:0}{
{\bs Y}_{00} =  \frac{1}{\sqrt{4\pi}} {\bs{\hat r}}\,,~~~
{\bs \Psi}_{00} = 0\,,~~~ {\bs \Phi}_{00} = 0 \,.
}

The first nontrivial vector spherical harmonics starts from the orbital momentum $l=1$:
\Align{App:Eq:VSH:1}{
      {\bs Y}_{10} {=} \phantom{-}\sqrt{\frac{3}{4 \pi}}\cos\theta\,{\bs{\hat r}},
&\quad{\bs Y}_{11} {=} - \sqrt{\frac{3}{8\pi}}e^{i \phi}\sin\theta\,{\bs{\hat r}}, 
\\
{\bs \Psi}_{10} {=} - \sqrt{\frac{3}{4 \pi}} \sin \theta \, {\bs{\hat \theta}}, 
& \quad {\bs \Psi}_{11} {=} - \sqrt{\frac{3}{8 \pi}} e^{i \phi}  
	\left(\cos \theta \,{\bs{\hat \theta}} + i {\bs{\hat \phi}}\right)
	,	\nonumber\\
{\bs \Phi}_{10} {=} - \sqrt{\frac{3}{4 \pi}} \sin \theta \, {\bs{\hat \phi}},
& \quad {\bs \Phi}_{11} {=} \phantom{-} \sqrt{\frac{3}{8 \pi}} e^{i \phi}  
	\left(i {\bs{\hat \theta}} - \cos\theta \, {\bs{\hat \phi}}\right)	
	.		\nonumber
}
\begin{widetext}
The basis for $l=2$ vector spherical functions is as follows:
\Align{App:Eq:VSH:2}{
& {\bs Y}_{20} = \phantom{-} \frac{1}{4} \sqrt{\frac{5}{\pi}} 
			\left( 3\cos^2\theta - 1 \right) \, {\bs{\hat r}},  
&& \quad {\bs Y}_{21} = - \sqrt{\frac{15}{8 \pi}} e^{i \phi} 
			\sin \theta \cos\theta \, {\bs{\hat r}}, 
&& {\bs Y}_{22} = \frac{1}{4}\sqrt{\frac{15}{2 \pi}} e^{2 i \phi} 
			\sin^2 \theta \, {\bs{\hat r}}, 
\\
& {\bs \Psi}_{20} = - \frac{3}{2} \sqrt{\frac{5}{\pi}} 
			\sin\theta \cos\theta \, {\bs{\hat \theta}}, 
&& \quad {\bs \Psi}_{21} = - \sqrt{\frac{15}{8 \pi}} e^{i \phi}  
			\left(\cos2\theta\,{\bs{\hat\theta}} + i\cos\theta{\bs{\hat \phi}}\right), 
&& {\bs \Psi}_{22} =  \sqrt{\frac{15}{8 \pi}} e^{2 i \phi} 
			\sin\theta\left(\cos\theta\,{\bs{\hat\theta}}+i{\bs{\hat\phi}}\right), 
\nonumber\\
& {\bs \Phi}_{20} = - \frac{3}{2} \sqrt{\frac{5}{\pi}} 
			\sin\theta \cos\theta \, {\bs{\hat \phi}},
&& \quad {\bs \Phi}_{21} = \phantom{-} \sqrt{\frac{15}{8 \pi}} e^{i \phi}  
			\left(i\cos\theta{\bs{\hat\theta}}-\cos2\theta\,{\bs{\hat\phi}}\right),
&& {\bs \Phi}_{22} =  \sqrt{\frac{15}{8 \pi}} e^{2 i \phi} 
			\sin\theta\left(-i{\bs{\hat\theta}}+\cos\theta\,{\bs{\hat \phi}}\right),
\nonumber
}

\subsection{Integrals of spherical functions} 
\label{App:Sec:VSH:itegrals}

The nonvanishing integrals are as follows:
\begin{align}
    \int d \Omega\,  {\bs Y}_{10} & = \frac{1}{2} \int d \Omega\, {\bs \Psi}_{10}  
                                    = \phantom{-} \sqrt{\frac{4 \pi}{3}} \hat {\bs z}\,, 
\qquad\qquad 
%
    \int d \Omega {\bs Y}_{11} = \frac{1}{2} \int d \Omega\, {\bs \Psi}_{11}  
                                 = - \sqrt{\frac{2 \pi}{3}} \bigl(\hat {\bs x} + i \hat {\bs y}\bigr)\,.
\end{align}

\end{widetext}


%

\end{document}